\documentclass[11pt]{article}

\usepackage[]{graphics,graphicx}

\usepackage{pstricks}

\usepackage{blindtext}

\newcommand{\Pd}[2]{\frac{\partial #1}{\partial #2}}
\newcommand{\lb}{\left(}
\newcommand{\rb}{\right)}
\newcommand{\D}{\displaystyle}

\usepackage{setspace} 

\RequirePackage{lineno}

\begin{document} 

 \modulolinenumbers[1]


\begin{center}
{{\LARGE\bf A mechanical erosion model for two-phase mass flows}}
\\[5mm]
{Shiva P. Pudasaini
\\[1mm]
{Department of Geophysics, Steinmann Institute, University of Bonn}\\
{Meckenheimer Allee 176, D-53115, Bonn, Germany}\\[1mm]
{E-mail: pudasaini@geo.uni-bonn.de}\\[5mm]
Jan-Thomas Fischer\\[1mm]
Austrian Research Centre for Forests, Department of Natural Hazards\\
Rennweg 1, A-6020, Innsbruck, Austria
\\[10mm]
}
\end{center}
\noindent
{\bf Abstract:} 
Erosion, entrainment and deposition are complex and dominant, but yet poorly understood, mechanical processes in geophysical mass flows. Here, we propose a novel, process-based, two-phase, erosion-deposition model capable of adequately describing these complex phenomena commonly observed in landslides, avalanches, debris flows and bedload transport. The model is based on the jump in the momentum flux including changes of material and flow properties along the flow-bed interface and enhances an existing general two-phase mass flow model (Pudasaini, 2012). A two-phase variably saturated erodible basal morphology is introduced and allows for the evolution of erosion-deposition-depths, incorporating the inherent physical process including momentum and rheological changes of the flowing mixture. By rigorous derivation, we show that appropriate incorporation of the mass and momentum productions or losses in conservative model formulation is essential for the physically correct and mathematically consistent description of erosion-entrainment-deposition processes. We show that mechanically deposition is the reversed process of erosion. We derive mechanically consistent closures for coefficients emerging in the erosion rate models. We prove that effectively reduced friction in erosion is equivalent to the momentum production. With this, we solve the long standing dilemma of mass mobility, and show that erosion enhances the mass flow mobility. The novel enhanced real two-phase model reveals some major aspects of the mechanics associated with erosion, entrainment and deposition. The model appropriately captures the emergence and propagation of complex frontal surge dynamics associated with the frontal ambient-drag with erosion.

\section{Introduction}

Erosion, entrainment and deposition processes play important role in debris flow dynamics and deposition morphology, and shaping the landscape. However, these processes are very poorly understood. Debris flows are gravity driven mixture flows of soil, rock, and water (Berger et al., 2011; McCoy et al., 2012) that can be modelled as a two-phase mixture flow of viscous fluid and solid particles (Pudasaini, 2012). Debris flows can dramatically increase their volume and destructive potential, and become exceptionally mobile by entraining sediment by scouring channel beds or undermining banks (Hungr et al., 2005; Reid et al., 2011). Entrainment can strongly influence the flow dynamics and the characteristics of deposit with adverse societal and environmental impacts (Rickenmann, 2005; Godt and Coe, 2007; Berger et al., 2011; Pirulli and Pastor, 2012). Different field and laboratory studies have investigated bed sediment entrainment (Egashira et al., 2001; Rickenmann et al., 2003; Hungr and Evans, 2004; Berger et al., 2011; Reid et al., 2011; Iverson, 2012; McCoy et al., 2012). Several mechanical and numerical models have included entrainment (Brufau et al., 2000; Fraccarollo and Capart, 2002; McDougall and Hungr, 2005; Chen et al., 2006; Mangeney et al., 2007; Bouchut et al., 2008; Tai and Kuo, 2008; Armanini et al., 2009; Crosta et al., 2009; Le and Pitman, 2009; Iverson, 2012; Fischer et al., 2015). Erosion may depend on the flow depth, flow velocity, solid concentration, density ratio, bed slope or, the effective stresses at the interface, and initial and boundary conditions (Gauer and Issler, 2004; Fagents and Baloga, 2006; Sovilla et al., 2006; Issler et al., 2008; Crosta et al., 2009; Mangeney et al., 2010; Berger et al., 2011).
\\[3mm]
There exist several hypotheses explaining the possible bed erosion mechanics. However, how, where and when the erodible material enters the moving mass is still a largely unanswered question, and poses significant modeling and computational challenges. Another critical aspect, almost not explicitly yet considered in a mechanically meaningful way is the deposition process. Although it may seem to be the opposite to erosion, there can be fundamental differences between erosion and deposition process.
\\[3mm]
Basically two types of erosion (and deposition) models are in use: empirical and mechanical ones. Empirical laws are developed with experience and are most often used in practice. They are based on yield- or erosion rates as user specified calibration parameters (Rickenmann et al., 2003; McDougall and Hungr, 2005; Chen et al., 2006) in terms of erosion volume, mean shear stress, erosion area and travel distance. Erosion rate models for debris flows also include equilibrium concentrations or slopes (Takahashi and Kuang, 1986; Egashira and Ashida, 1987; Takahashi et al., 1992;  Egashira, 1993; Ghilardi and Natale, 1998; Brufau et al., 2000; Egashira et al., 2001; Chen and Zhang, 2015). Tai and Kuo (2008) employed moving coordinates to generate erodible bed. Assuming small velocity, Bouchut et al. (2008) obtained evolving interface that includes basal curvature and erosion rate. Extending Takahashi and Kuang (1986), Le and Pitman (2009) obtained erosion rate as a linear combination of velocity and shear stress thresholds with basal velocity.
\\[3mm]
Process-based mechanical models are derived with the mass and momentum exchanges between a debris flow and the underlying erodible bed. This results in the erosion rate that is proportional to the shear stress difference between entraining and resisting stresses; and inversely proportional to the effective slip velocity on either side of the bed (Fraccarollo and Capart, 2002; Iverson, 2012; Issler, 2014). However, this can simply be obtained directly by considering the balance between the involved forces and momentum fluxes during the erosion process. The solid-like shear stress along the substrate is often closed by effective Coulomb friction. In Fraccarollo and Capart (2002) the fluid-like interface shear stress along the moving debris is closed with the Chezy equation. In contrast, Iverson (2012) assumed complete liquefaction of the substrate and Coulomb-friction for the sliding layer. However, it appears that the mechanical erosion rate introduces a singularity, such as the erosion-rate is inversely proportional to the velocity. For bulk-, or solid-type models (Iverson, 2012) as flow slows down or stops, erosion-rate becomes infinitely large, and as flow moves with very high speed, erosion rate is negligible, which differs from our intuition and observations in nature.
\\[3mm]
 Erosion models can also be categorized as single-phase or mixture models. Most erosion models are developed for single-phase viscous fluid including Chezy-type equation, or viscous boundary (Hogg and Pritchard, 2004). Similarly, erosion rates are used commonly for landslides, rock and debris avalanches (Naaim et al., 2003; McDougall and Hungr, 2005; Bouchut et al., 2008; Tai and Kuo, 2008; Le and Pitman, 2009; Mangeney et al., 2010). One of the early mixture simulations with erosion were presented by Armanini et al. (2009), but for effectively single-phase material. McDougall and Hungr (2005), Crosta et al. (2009), and Pirulli and Pastor (2012) presented one of the very first simulations for rock/debris-avalanches with entrainment/deposition. Very few erosion models are developed for fluid-grain mixture (Fraccarollo and Capart, 2002), saturated entrained materials (Crosta et al., 2009), and debris mixture (Armanini et al., 2009; Iverson, 2012). None of these models is fully coupled and true two-phase model. Only very few erosion rate equations are used so far for simulation (Pirulli and Pastor, 2012). 
\\[3mm]
 Another prevailing aspect, is that, all these models are based on the bulk mixture and are effectively single-phase. However, only two-phase erosion models would better describe the phenomena as the debris flow and the erodible substrate themselves typically are two-phase materials. Importantly, realistic two-phase erosion, entrainment and deposition models can only be constructed by considering the two-phase mass flow model that explicitly considers both the solid and fluid phases and the strong interactions between the phases including the drag and viscous effects (Pudasaini, 2012). True two-phase erosion/deposition models, taking into account both the solid and fluid phases separately and explicitly with possible phase interactions, have not yet been developed. Erosion models applicable to real two-phase flows have recently been introduced by Pudasaini and Fischer (2016). 
\\[3mm]
Despite the importance of entrainment to hazard assessment and landscape evolution, clear understanding of the basic process still remains elusive owing to a lack of high-resolution, field-scale data, and also laboratory experiments are limited to few flow parameters. Physics-based numerical simulations may overcome these limitations and facilitate for the more complete understanding by investigating much wider aspects of the flow parameters, erosion, mobility and deposition. Although in the recent years different experimental and theoretical works, and simulations have focused on sediment entrainment, quantitative and mechanical constraints on rates and forms are still limited and no consensus has been reached yet (Bouchut et al., 2008; Luca et al., 2009; Mangeney et al., 2010; Iverson, 2012; McCoy et al., 2012). The mixture composition can evolve to dramatically change the spatial distribution of frictional and viscous resistance in bulk material and the boundaries (Iverson et al., 2011; Pirulli and Pastor, 2012; Pudasaini and Krautblatter, 2014). So, it is very important to properly model two-phase bed and flow properties that strongly control the occurrence and rates of entrainment, and mobility.
\\[3mm]
However, even the basic mechanism of erosion in the presence of both fluid and solid has not yet been touched for which more advanced models are necessary, including mass and momentum productions in the respective balance equations for phases. For solid, additional contribution emerges from basal erosion, and deposition. For fluid, different contributions are due to stream falling on debris, and water run-off leaving the debris. Erosion/deposition processes will have apparent effects with respect to a real two-phase, rather than in effectively single-phase flow. The immediate impacts are seen in: the dynamics of the solid and fluid volume fractions, and changes in the solid and fluid densities. These largely influence the interfacial drag, and virtual mass forces. As the basal surface changes effective internal and basal frictions evolve due to solid and fluid entrainment.
\\[3mm]
Spatially changing fluid property may substantially alter the fluid viscosity. Spatially changing amount of solid influences the non-Newtonian fluid stress. Evolving contrast between the solid and fluid density leads to change in buoyancy. This plus evolving earth pressure alters the solid pressure. Erosion/deposition changes the basal surface and the whole system through the Coulomb friction and drag terms.  These aspects have (almost) never been investigated as they are related to the true two-phase nature of flow. In the computational models by Fraccarollo and Capart (2002) and Armanini et al. (2009), only the mass balance equations contain the bed elevation, but the momentum equations did not. Changes in the basal surface due to erosion have only recently been included in Fraccarollo and Capart (2002) and Le and Pitman (2009). The process on how the entrained mass is accelerated and distributed within the flow and how it decelerates and deposits can be simulated with a real two-phase mass flow and erosion-deposition model. 
\\[3mm]
From mechanical and mathematical point of view there are five major aspects in erosion (and/or deposition) modelling in geophysical mass flows that potentially also apply to other flow types. The first is the erosion-rate. Due to the complex flow dynamics and the rheology of the flowing mixture, the composition and state of the (erodible) basal morphology and its dynamic response, proper understanding of the erosion process is, perhaps, one of the most challenging tasks in geophysical mass flows. This is, because erosion results from two competitive forces: (i) the force exerted by the moving mixture on the erodible bed, and (ii) the resistance by later on the moving material. Proper description of these forces involves their fundamental mechanical behaviors and how these are changing during the flow and interactions with the mobile substrate. There are well known methods to describe the basic mechanics of both the moving material and the erodible bed, if their mechanics do not change in time and/or space. Nevertheless, as the physics of the moving mixture material may change largely during the flow, and the same may be true for the flow bed, understanding and modelling the erosion rates is a great challenge. Second, how to model the real erosion process is another crucial aspect. Because, often in the literature only effectively single-phase mass movements and the associated single-phase erosion is considered. Nevertheless, in reality, the debris flow material and the erodible substrate both are most often composed of the two-phase solid sediment particles and the interstitial viscous fluid. Third, the proper knowledge of the velocities of the fluid and solid particles that have just been eroded around the interface, and at the lowest layer of the flow, is of major importance. There is no trivial way to estimate this. Nevertheless, this plays a very important role in the entire erosion and transport process. Often, the velocity of eroded material is set to zero that results in zero momentum productions leading to completely different scenario. Since the eroded material moves with non-zero velocity, setting this velocity to zero is physically not admissible. The fourth aspect concerns the momentum productions (or, losses) resulting from the mass productions (or, losses). Two different model-types are available even for the effectively single-phase and quasi two-phase mass flow modelling related to the momentum production. One type of models include the momentum production (Gray, 2001; Pudasaini and Hutter, 2007; Le and Pitman, 2009) whereas the other model types consider the mass productions but do not include, or argue not to include, the corresponding momentum production or loss in the momentum balance equations (Christen et al., 2010; Iverson, 2012; Pirulli and Pastor, 2012). However, the rigorous mathematical derivation clearly shows that the conservative formulation, as presented here, demands for the inclusion of the momentum productions in the momentum balance equations. Otherwise, those models are physically incorrect and mathematically inconsistent. The fifth aspect is concerned about the erosion induced mobility, which still is a dilemma, as there are two conflicting arguments concerning enhanced or decreased mobility due to erosion.
\\[3mm]
This paper addresses these aspects and presents a novel mechanical two-phase erosion model for geophysical mass flows such as debris flows, and particle-fluid transports on mountain slopes and channels. The new model is based the general two-phase mass flow model (Pudasaini, 2012), and on the jump in the momentum flux across the erodible interface (Drew, 1983) where the surface tension has been neglected. The formally derived model equations are in conservative form and consistently include both mass and momentum production. We prove that in erosional settings the effectively reduced frictional stress is equivalent to the (resulting) momentum production. We solve the long standing dilemma of mass mobility, and show that erosion enhances the flow mobility. The simulations reveal some major and novel mechanical aspects associated with erosion, entrainment and deposition.

\section{Two-phase mass flow model with erosion-deposition}
    
{In two-phase debris mixtures, phases are characterized by different material properties. The fluid phase is characterized by its material density $\rho_{f}$, viscosity $\eta_{f}$  and isotropic stress distribution; whereas the solid phase is characterized by its material density $\rho_{s}$, the internal friction angle $\phi$, the basal friction angle $\delta$, an anisotropic stress distribution, and the lateral earth pressure coefficient $K$. The subscripts $s$ and $f$ represent the solid and the fluid phases respectively, with the depth-averaged velocity components for fluid  $\textbf{u}_{f}$ = ($u_{f}$, $v_{f}$) and for solid $\textbf{u}_{s}$ = ($u_{s}$, $v_{s}$) in the down-slope $(x)$ and the cross-slope $(y)$ directions. The total flow depth is denoted by $h$, and the solid volume fraction $\alpha_s$ (similarly the fluid volume fraction $\alpha_f = 1- \alpha_s$) are functions of space and time. The solid and fluid mass balance equations  (Pudasaini, 2012) together with the evolution equation for the basal morphology are given by}
\begin{eqnarray}
\begin{array}{lll}
\D{\Pd{}{t}{\lb \alpha_s h\rb} + \frac{\partial}{\partial x}{\lb \alpha_s h u_s\rb}
                 + \frac{\partial}{\partial y}{\lb \alpha_s h v_s\rb}=E_s},
\\[5mm]
\D{\Pd{}{t}{\lb \alpha_f h\rb} + \frac{\partial}{\partial x}{\lb \alpha_f h u_f\rb}
                 + \frac{\partial}{\partial y}{\lb \alpha_f h v_f\rb}=E_f,}
\\[5mm]
\D{\Pd{b}{t}= -E; \,\,\,\,\,\,\, E = E_s + E_f,}
\end{array}    
\label{Model_Final_Mass}
\end{eqnarray}
where $b = b(x,y; t)$ is the basal topography that evolves in space and time, and $E_s$, $E_f$ are the solid and the fluid erosion rates, and $E$ is the total erosion rate. This model can be used for partially or, fully saturated erodible basal substrate, or the substrate that is not erodible ($E = 0$). When the basal substrate is erodible, the solid fraction of $E$, i.e., $E_s$, enters into the solid mass balance as the solid mass production. So does the fluid fraction of $E$, i.e., $E_f$, that enters into the fluid mass balance as the fluid mass production. Models for these erosion rates are developed in Section 3.
\\[3mm]
 Similarly, momentum conservation equations for the solid and the fluid phases, respectively, are: 
\begin{eqnarray}
\begin{array}{lll}
\resizebox{.935\hsize}{!}{$\D{\Pd{}{t}\biggl [ \alpha_s h \lb u_s \!-\! \gamma \mathcal C\lb u_f\! -\!u_s \rb \rb \biggr ]
  \!+\!\Pd{}{x}\biggl [ \alpha_s h \lb u_s^2 \!-\! \gamma \mathcal C\lb u_f^2 \!-\!u_s^2 \rb\!+\! \beta_{x_s} \frac{h}{2}\rb \biggr ]
  \!+\!\Pd{}{y}\biggl[ \alpha_s h \lb u_sv_s \!-\! \gamma \mathcal C\lb u_fv_f \!-\!u_sv_s \rb \rb \biggr ]}
\D{=  h\mathcal S_{x_s} \!+\! u_s^bE_s}$},\\[5mm]  
\resizebox{.935\hsize}{!}{$\D{\Pd{}{t}\biggl [ \alpha_s h \lb v_s \!-\! \gamma \mathcal C\lb v_f \!-\!v_s \rb \rb \biggr ]
  \!+\!\Pd{}{x}\biggl [ \alpha_s h \lb u_sv_s \!-\! \gamma \mathcal C\lb u_fv_f \!-\!u_sv_s \rb\rb \biggr ]
  \!+\!\Pd{}{y}\left[ \alpha_s h \lb v_s^2 \!-\! \gamma \mathcal C\lb v_f^2 \!-\!v_s^2\rb\!+\! \beta_{y_s} \frac{h}{2}  \rb \right ]}
\D{=  h\mathcal S_{y_s}\!+\! v_s^bE_s}$},\\[5mm]
\resizebox{.935\hsize}{!}{$ \D{\Pd{}{t}\left [ \alpha_f h \lb u_f \!+\! \frac{\alpha_s }{\alpha_f}\mathcal C\lb u_f \!-\!u_s \rb \rb \right ]
  \!+\!\Pd{}{x}\left [ \alpha_f h \lb u_f^2 \!+\! \frac{\alpha_s }{\alpha_f}\mathcal C\lb u_f^2 \!-\!u_s^2 \rb  \!+\! \beta_{x_f} \frac{h}{2}\rb \right ]
  \!+\!\Pd{}{y}\left[ \alpha_f h \lb u_fv_f  \!+\! \frac{\alpha}{\alpha_f}\mathcal C\lb u_fv_f \!-\!u_sv_s \rb \rb \right ]
=  h\mathcal S_{x_f}\!+\! u_f^bE_f}$},\\[5mm] 
\resizebox{.935\hsize}{!}{$ \D{\Pd{}{t}\left [ \alpha_f h \lb v_f \!+\! \frac{\alpha_s }{\alpha_f}\mathcal C\lb v_f \!-\!v_s \rb \rb \right ]
 \! +\!\Pd{}{x}\left [ \alpha_f h \lb u_fv_f \!+\! \frac{\alpha_s }{\alpha_f}\mathcal C\lb u_fv_f \!-\!u_sv_s \rb\rb \right ]
  \!+\!\Pd{}{y}\left[ \alpha_f h \lb v_f^2  \!+\! \frac{\alpha_s}{\alpha_f}\mathcal C\lb v_f^2 \!-\!v_s^2 \rb \!+\!  \beta_{y_f} \frac{h}{2}\rb \right ]
=  h\mathcal S_{y_f}\!+\! v_f^bE_f}$}.
\end{array}    
\label{Model_Final}
\end{eqnarray}
These solid and fluid momentum equations are rigorously derived (Pudasaini, 2012) and include the solid and fluid momentum production terms, second terms on the right hand sides. One of the important aspects in these momentum production terms are, that the velocities of the solid and the fluid particles at the bottom that have just been eroded, i.e., ($u_s^b, v_s^b; u_f^b, v_f^b$) are different than the depth-averaged (mean) velocities ($u_s, v_s; u_f, v_f$) that appear in the inertial (or, the convective) part, and also the source terms, of the mass and momentum equations. In (\ref{Model_Final}), the source terms are as follows:
\begin{eqnarray}
\resizebox{.935\hsize}{!}{$
\mathcal S_{x_s}\! = \alpha_s\left [g^x \!- \frac{u_s}{|{\bf u}_s|}\tan\delta p_{b_s} \!-\varepsilon p_{b_s}\Pd{b}{x}\right ] 
\!-\!\varepsilon \alpha_s\gamma p_{b_f}\!\left [ \Pd{h}{x} + \Pd{b}{x}\right ]
\!+ C_{DG} \lb u_f - u_s \rb{ |{\bf u}_f - {\bf u}_s|}^{\jmath-1} 
\!-C_{DV}^{x_s} u_s|{\bf u}_s|^{\jmath-1} \alpha_s,$}
\label{Model_Final_ss}\\[5mm]
\resizebox{.935\hsize}{!}{$\mathcal S_{y_s} \!= \alpha_s\left [ g^y \!- \frac{v_s}{|{\bf u}_s|}\tan\delta p_{b_s} \!-\varepsilon p_{b_s}\Pd{b}{y}\right ] 
\!-\!\varepsilon \alpha_s\gamma p_{b_f}\!\left [ \Pd{h}{y} + \Pd{b}{y}\right ]
\!+ C_{DG} \lb v_f - v_s \rb{ |{\bf u}_f - {\bf u}_s|}^{\jmath-1}
\!-C_{DV}^{y_s} v_s|{\bf u}_s|^{\jmath-1} \alpha_s,$}
\label{Model_Final_s}
\end{eqnarray}
\vspace{-3mm}
\begin{eqnarray}
\begin{array}{lll}
\D{\mathcal S_{x_f} = \alpha_f\biggl [g^x - \varepsilon   \biggl [\frac{1}{2}p_{b_f}\frac{h}{\alpha_f}\Pd{\alpha_s}{x} +  p_{b_f}\Pd{b}{x}
 -\frac{1}{\alpha_fN_R}\left \{ 2\frac{\partial^2 u_f}{\partial x^2}+  \frac{\partial^2v_f}{\partial y\partial x}
 + \frac{\partial^2 u_f}{\partial y^2} - \frac{\chi u_f}{\varepsilon^2h^2} \right \}} \\[5mm]
 +\D{ \frac{1}{\alpha_fN_{R_\mathcal A}}\left \{ 2\Pd{}{x}\lb \Pd{\alpha_s}{x}\lb u_f - u_s\rb\rb 
 + \Pd{}{y}\lb \Pd{\alpha_s}{x}\lb v_f -v_s\rb + \Pd{\alpha_s}{y}\lb u_f - u_s\rb\rb\right\}
-\frac{\xi\alpha_s\lb u_f -u_s\rb}{\varepsilon^2 \alpha_fN_{R_\mathcal A}h^2} 
 \biggr]\biggl ]}\\[5mm] 
-\D{\frac{1}{\gamma}C_{DG}\lb u_f - u_s \rb{ |{\bf u}_f - {\bf u}_s|}^{\jmath-1}
 -C_{DV}^{x_f} u_f|{\bf u}_f|^{\jmath-1} \alpha_f},
\end{array}    
\label{Model_Final_fx}
\end{eqnarray}
\vspace{-3mm}
\begin{eqnarray}
\begin{array}{lll}
\D{\mathcal S_{y_f} = \alpha_f\biggl [g^y - \varepsilon   \biggl [ \frac{1}{2}p_{b_f}\frac{h}{\alpha_f}\Pd{\alpha_s}{y}+  p_{b_f}\Pd{b}{y}
 -\frac{1}{\alpha_fN_R}\left \{ 2\frac{\partial^2 v_f}{\partial y^2}+  \frac{\partial^2u_f}{\partial x\partial y}
 + \frac{\partial^2 v_f}{\partial x^2} - \frac{\chi v_f}{\varepsilon^2h^2} \right \}} \\[5mm]
 + \D{\frac{1}{\alpha_fN_{R_\mathcal A}}\left \{ 2\Pd{}{y}\lb \Pd{\alpha_s}{y}\lb v_f - v_s\rb\rb 
 + \Pd{}{x}\lb \Pd{\alpha_s}{y}\lb u_f -u_s\rb + \Pd{\alpha_s}{x}\lb v_f - v_s\rb\rb\right\}
-\frac{\xi \alpha_s\lb v_f -v_s\rb}{\varepsilon^2 \alpha_fN_{R_\mathcal A}h^2} 
 \biggr]\biggl ]}\\[5mm] 
-\D{\frac{1}{\gamma}C_{DG}\lb v_f - v_s \rb{ |{\bf u}_f - {\bf u}_s|}^{\jmath-1}
-C_{DV}^{y_f} v_f|{\bf u}_f|^{\jmath-1} \alpha_f}.
\end{array}    
\label{Model_Final_fy}
\end{eqnarray}
The pressures and the other parameters involved in the above model equations are as follows:
\begin{eqnarray}
\begin{array}{lll}
\D{\beta_{x_s} = \varepsilon K_x p_{b_s}, \,\,\,\, \beta_{y_s} = \varepsilon K_y p_{b_s},\,\,\,\beta_{x_f} = \beta_{y_f} = \varepsilon p_{b_f},\,\,\, p_{b_f} = - g^z, \,\,\, p_{b_s} = (1-\gamma)p_{b_f},}\\[5mm]
\D{C_{DG} = \frac{\alpha_s \alpha_f(1-\gamma)}{\left [\varepsilon \mathcal U_T\{{\cal P}\mathcal F(Re_p) + (1-{\cal P})\mathcal G(Re_p)\}\right ]^{\jmath}},\,\,\,\,
\mathcal F = \frac{\gamma}{180}\lb\frac{\alpha_f}{\alpha_s} \rb^3 Re_p, \,\,\,\,  \mathcal G= \alpha_f^{M(Re_p) -1},}
\\[5mm]
\D{\gamma =\frac{\rho_f}{\rho_s},\, Re_p = \frac{\rho_f d~ \mathcal U_T}{\eta_f},\, N_R = \frac{\sqrt{gL}H\rho_f}{\alpha_f\eta_f}, N_{R_{\mathcal A}} = \frac{\sqrt{gL}H\rho_f}{\mathcal A \eta_f},\,
\alpha_f = 1-\alpha_s,\, \mathcal A = \mathcal A(\alpha_f).}
\end{array}    
\label{Model_Final_parameters}
\end{eqnarray}
 Equations (\ref{Model_Final_Mass}) are the depth-averaged mass balances for solid and fluid phases respectively, and (\ref{Model_Final}) are the depth-averaged momentum balances for solid (first two equations) and fluid (other two equations) in the $x$- and $y$-directions, respectively.
\\[3mm]
In the above {non-dimensional equations (\ref{Model_Final_Mass})-(\ref{Model_Final}), $x$, $y$ and $z$ are the locally orthogonal coordinates in} the down-slope, cross-slope and flow normal directions, and $g^x$, $g^y$, $g^z$ are the respective components of gravitational acceleration. $L$ and $H$ are the typical length and depth of the flow, $\varepsilon = {H}/{L}$ is the aspect ratio, and $\mu =\tan\delta$ is the basal friction coefficient. $C_{DG}$ is the generalized drag coefficient. Simple linear (laminar-type, at low velocity) or quadratic (turbulent-type, at high velocity) drag is associated with ${\jmath} = 1$ or $2$, respectively. $\mathcal{U}_{T}$ is the terminal velocity of a particle and $\mathcal{P}\in [0,1]$ is a parameter which combines the solid-like ($\mathcal{G}$) and fluid-like ($\mathcal{F}$) drag contributions to flow resistance. $p_{b_{f}}$ and $p_{b_{s}}$ are the effective fluid and solid pressures. $\gamma$ is the density ratio, $\mathcal{C}$ is the virtual mass coefficient (kinetic energy of the fluid phase induced by solid particles), $M$ is a function of the particle Reynolds number ($R_{e_{p}}$), $\chi$ includes vertical shearing of fluid velocity, and $\xi$ takes into account different distributions of $\alpha_s$. $\mathcal{A}$ is the mobility of the fluid at the interface, and $N_{R}$ and $N_{R_{\mathcal{A}}}$, respectively, are the quasi-Reynolds number and mobility-Reynolds number associated with the classical Newtonian and enhanced non-Newtonian fluid viscous stresses. $C_{DV}$ are the ambient viscous drag coefficients (Kattel et al., 2016). 
\\[3mm]
 The evolution of basal topography ${\partial b}/{\partial t}= -E$ in (\ref{Model_Final_Mass}) due to erosion and deposition is explicitly included in the model. With this, the basal change directly influences the source terms in (\ref{Model_Final_ss}) - (\ref{Model_Final_fy}) by accounting for changes that are associated with the driving forces in the net force balance. This appears very important for geophysical mass flows which are mainly driven by gravity and slope changes, i.e., the respective components of gravitational accelerations, Coulomb frictions, the basal and hydraulic pressure gradients, and the buoyancy induced terms.

\section{A  two-phase, process-based, non-singular mechanical model for erosion} 

As mentioned in the introduction, two different approaches to model the erosion rates are present in the literature, both of them (mostly) effectively single-phase. The first is the empirical and the second is the mechanical model. Empirical laws are relatively simple and only involve the overall bulk dynamical quantities, namely the flow depth and/or the bulk velocity coupled with an empirical erosion factor, $E = E\lb E^{ri}_{emp}, h u\rb$. Whereas the mechanical erosion models involve the dynamical variables associated with the flow coupled to several mechanical parameters of the flowing material and the erodible bed, $E = E\lb E^{ri}_{mech}, h/u\rb$. $E_{emp}^{ri}$ and $E_{mech}^{ri}$ are the empirical and mechanical erosion rate intensities ($^{ri}$-factors). Additionally, the evolving basal topography may lead to substantial changes, e.g., in accelerating and retarding gravitational components, and normal and shear loads. 
\\[3mm]
One of the major concerns related to erosion is understanding the process and describing it by mechanical models. Here, we develop a two-phase, process-based, non-singular mechanical model for erosion-deposition. The non-singularity refers to the finite and physically meaningful values of the erosion rates at the flow-bed interface. The mechanical erosion rate equations can be derived by simply considering the force balance at the interface between the cascading debris material and the erodible bed beneath it, and the momentum flux across the interface (Drew, 1983). Jumps in these quantities govern the erosion-rates. We develop the erosion rate models for both the solid and the fluid phases. 

\subsection{Solid erosion-rate}

First, we deal with the erosion rate for the solid phase. We consider the stresses on both sides of the erosion interface. The sliding mass applies the shear stress  $\tau_{s_l}^m$ (along $x$, the positive down slope flow direction) on the erodible bed, and the bed applies the shear stress $-\tau_s^b$ (opposite to the flow, in negative direction) against the flowing material. Here, $m$ stands for the moving debris mixture, $l$ for the lower layer of the moving mixture (so, on the upper side of the interface), and $b$ the erodible bed (so, on the lower side of the interface), respectively. Therefore, the resulting shear stress along the singular erodible interface is $\tau_{s_l}^m - \tau_{s}^b$. This is the shear stress jump across the interface, or the net shear stress in the system. Here, the singular interface refers to the surface for which the material properties and/or the dynamical quantities are different on the either side of this surface, i.e., a jump prevails in the relevant quantities across the interface. 
\\[3mm]
Next, we deal with the momentum flux to the sliding debris, and out of the debris generated by the erosion of the bed. Both the debris and eroded mass flow in the positive flow direction. The $x$-directional solid mass flux (in the debris  mixture) on the upper side of the interface is $\rho_s^m u_{s_l}^m\alpha_s^m$ where $\rho_s^m$ is the density of the solid in the mixture, $\alpha_s^m$ is the volume fraction, and $u_{s_l}^m$ is the velocity of the solid particle at the lowest layer in the mixture (i.e., on the upper side of the interface). Similarly, when erosion takes place, the $x$-directional solid mass flux in and of the eroded material is given by $\rho_s^b u_{s}^b \alpha_s^b$ where $b$ denotes the quantities associated with the erodible layer. So, the mass fluxes $\rho_s^m u_{s_l}^m\alpha_s^m$ and $\rho_s^b u_s^b\alpha_s^b$ are positive. As erosion takes place, the interface between these two mass fluxes moves in flow perpendicular direction (along $z$-direction) into the erodible bed. The speed of the singular interface between the moving material and the erodible basal layer is the erosion rate $E_s$ (erosion-velocity). The momentum flux (associated to the eroded material) from bed to the flowing material introduces a propagation of the erosion interface, where $E_s\rho_s^m u_{s_l}^m\alpha_s^m$ is the momentum flow into the moving debris. In the meanwhile, due to entrainment, the eroded debris material in the basal surface loses the mass, and thus, the flow normal momentum flux for the eroded mass is negative, $-E_s\rho_s^b u_s^b\alpha_s^b$. This is the amount by which the momentum is transferred from the eroded basal debris material. 
\\[3mm]
Erosion is the result of the net shear stress, and the momentum fluxes are induced by the applied shear stresses. Hence, the effectively reduced frictional net shear stress must be balanced by the induced net momentum flux: 
\begin{eqnarray}
\begin{array}{lll}
\D{\tau_{s_l}^m -\tau_s^b 
= E_s\lb \rho_s^m u_{s_l}^m \alpha_s^m-  \rho_s^b u_{s}^b\alpha_s^b\rb}.
\end{array}    
\label{MassProduction_Rate_a}
\end{eqnarray}
 This shows that the balance between the net stress and the net momentum flux results in the propagation of the erosion induced singular surface at the velocity $E_s$. The jump in the shear stress and the jump in the normal momentum fluxes must be balanced and can be written as a single jump in terms of the momentum flux: $||\tau_{s} - E_s\rho_s u_s\alpha_s||_-^+ = 0$, where $+, -$ stands for base of the sliding debris and top of the erodible substrate, respectively, and $E_s^+ = E_s^- = E_s$. This relation can also be derived by applying the momentum jump (Drew, 1983), or the Rankine-Hugoniot jump condition (Fraccarollo and Capart, 2002). Model as in (\ref{MassProduction_Rate_a}) has also been presented in Iverson (2012) by considering debris bulk momentum equations on either side of the interface.
\\[3mm]
One of the major tasks modeling the erosion rate $E_s$ is concerned with the shear stresses. The granular materials satisfy the Coulomb plastic strength (Savage and Hutter, 1989; Iverson, 1979; Pitman and Le, 2005; Pudasaini and Hutter, 2007). So, the shear stresses are described by the Coulomb law, and the net shear stress at the interface is given by (Pudasaini, 2012):
\begin{eqnarray}
\begin{array}{lll}
\D{\tau_{s_l}^m -\tau_s^b 
= \lb 1-\gamma^m\rb \rho_s^m~ g\cos\zeta h~\mu_s^m \alpha_s^m - \lb 1-\gamma^b\rb\rho_s^b~ g\cos\zeta h~\mu_s^b\alpha_s^b} \\[5mm]
\hspace{1.43cm}
= \D{g \cos\zeta h \left [\lb 1-\gamma^m\rb\rho_s^m \mu_s^m \alpha_s^m - \lb 1-\gamma^b\rb\rho_s^b \mu_s^b\alpha_s^b \right ]
},
\end{array}    
\label{MassProduction_Rate_b}
\end{eqnarray}
where $\zeta$ is the slope angle, $\lb \gamma^m = \rho_f^m/\rho_s^m, \gamma^b = \rho_f^b/\rho_s^b\rb$ are the fluid-solid density ratios, $\lb 1 - \gamma^m, 1-\gamma^b\rb$ are the buoyancy induced factors, ($\mu_s^m = \tan\delta_s^m, \mu_s^b = \tan\delta_s^b$) are the Coulomb friction coefficients, and ($\alpha_s^m, \alpha_s^b$) are the solid volume fractions on either sides of the interface. So, the erosion-rate can be expressed as
\begin{eqnarray}
\begin{array}{lll}
\D{E_s =\frac{g \cos\zeta h \left [ \lb 1-\gamma^m\rb\rho_s^m \mu_s^m \alpha_s^m - \lb 1-\gamma^b\rb\rho_s^b \mu_s^b\alpha_s^b \right ]}{\lb \rho_s^m u_{s_l}^m \alpha_s^m-  \rho_s^b u_{s}^b\alpha_s^b\rb }}\\[7mm]
\hspace{.55cm}
= \D{\frac{g \cos\zeta h \left [ \lb 1-\gamma^m\rb\rho_s^m \mu_s^m \alpha_s^m - \lb 1-\gamma^b\rb\rho_s^b \mu_s^b\alpha_s^b \right]}{\lb \rho_s^m \lambda_{s_l}^m \alpha_s^m-  \rho_s^b \lambda_{s}^b\alpha_s^b\rb u_s }
},
\end{array}    
\label{MassProduction_Rate_c}
\end{eqnarray}
where $\lambda_{s_l}^m$ and $\lambda_{s}^b$ are the erosion drift coefficients that connect the mean flow velocity $u_s$ to both the velocity of the flow at the lower level of the flow $u_{s_l}^m=\lambda_{s_l}^m\,u_s$, and the velocity of the eroded particles $u_{s}^b=\lambda_{s}^b\,u_s$. Here, $u_{s_l}^m$ and $u_{s}^b$ are the velocities of the (carrier) particles that are moving on either side of the interface under the influence of the debris flow velocity field $u_s$. Since $u_{s_l}^m$ and $u_{s}^b$ are associated with the different sides of the erosion interface the coefficients $\lambda_{s_l}^m$ and $\lambda_{s}^b$ are associated with the mobility, or the conductivity of the erosion drift. In the following, we refer to $\lambda_{s_l}^m$ and $\lambda_{s}^b$ as erosion drift coefficients. $u_s$ can further include the shape factor (Hutter et al., 2005; Christen et al., 2010; Castro-Orgaz et al., 2015) that can also be related to the erosion drift coefficients. The minimum of the lower bounds of the drift coefficients $\lambda_{s_l}^m$ and $\lambda_{s}^b$ may be zero, but not simultaneously, while the maximum of their upper bounds may be unity. However, usually, these are distinct positive quantities, and combined with the possibly different densities and volume fractions across the interface, result in a non-singular erosion rate (see, later). 
\\[3mm] 
{The jump in the momentum flux across the interface is induced by the shear stress jump. This is equivalent to the shear velocity (jump) associated with the jump in the relevant densities, volume fractions and velocities across the interface. With the parameters $\lambda_{s_l}^m$ and $\lambda_{s}^b$, $u_s$ constitutes a representative velocity induced by the shearing, because $\lambda_{s_l}^m, \lambda_{s}^b; u_{s_l}^m, u_{s}^b$ are the result of the shearing $\tau_{s_l}^m$ and $\tau_s^b$. So, $u_s$ is proportional to the shear velocity $u_*$ of the system which is given by the square root of the ratio between the net shear stress of the system $\tau_{s_l}^m -\tau_s^b$ and the relevant (or, representative) net density across the interface, $\rho_s^m \lambda_{s_l}^m \alpha_s^m-  \rho_s^b \lambda_{s}^b\alpha_s^b$. That is, $u_* = \sqrt{\lb \tau_{s_l}^m -\tau_s^b\rb{\Big /}\lb \rho_s^m \lambda_{s_l}^m \alpha_s^m-  \rho_s^b \lambda_{s}^b\alpha_s^b\rb}$. Since $u_s \propto u_*$, there exists a proportionality factor $\tilde \nu$ such that $u_s = \tilde \nu\,u_*$. Thus, $\D{u_s = \sqrt{\nu\lb \tau_{s_l}^m -\tau_s^b\rb}{\Big /}\sqrt{\lb \rho_s^m \lambda_{s_l}^m \alpha_s^m-  \rho_s^b \lambda_{s}^b\alpha_s^b\rb}}$, where $\tilde \nu = \sqrt{\nu}$ is set for simplicity.} The mechanical significance of $\tilde \nu $ is discussed later. Employing this, (\ref{MassProduction_Rate_c}) yields:
\begin{eqnarray}
\begin{array}{lll}
\D{E_s =
\frac{\sqrt{\lb 1-\gamma^m\rb\rho_s^m \mu_s^m \alpha_s^m - \lb 1-\gamma^b\rb\rho_s^b \mu_s^b\alpha_s^b\,}}
{\sqrt{\nu\lb \rho_s^m \lambda_{s_l}^m \alpha_s^m-  \rho_s^b \lambda_{s}^b\alpha_s^b\rb}} \sqrt{g \cos\zeta h}
\,\,= \,E_s^{ri} \sqrt{g\,\cos\zeta\,h}},
\end{array}    
\label{MassProduction_Rate_d}
\end{eqnarray}
where $E_s^{ri} = {\sqrt{\lb 1-\gamma^m\rb\rho_s^m \mu_s^m \alpha_s^m - \lb 1-\gamma^b\rb\rho_s^b \mu_s^b\alpha_s^b \,}}{\Big/}{\sqrt{\nu( \rho_s^m \lambda_{s_l}^m \alpha_s^m-  \rho_s^b \lambda_{s}^b\alpha_s^b)}}.$ The relation (\ref{MassProduction_Rate_d}) takes into account the inertia of the erodible bed material due to density (and the solid fraction) difference across the interface, i.e., the difference of these physical properties (and dynamical variables) between the flowing material and the bed material (mixture). 
\\[3mm]
The solid erosion rate (\ref{MassProduction_Rate_d}) states that triggering of erosion (or, beginning of the deposition) process collectively depends on the erosive capacity due to the shear stress exerted by the flowing mixture $\lb \lb 1-\gamma^m\rb\rho_s^m \mu_s^m \alpha_s^m\rb $ and the inertia or, the resistance of the bed material $\lb \lb 1-\gamma^b\rb\rho_s^b \mu_s^b\alpha_s^b\rb $. The effect of the fluid in the bed can be incorporated through the evolving fluid fraction, $\lb \alpha_f^b = 1 - \alpha_s^b\rb$ in the bed. For undrained conditions (e.g., clay material), evolving pore fluid pressure can be included through the effective friction coefficient, $\tilde \mu_s^b$. For the drained condition and large hydraulic conductivity, e.g., the debris composed of gravel, the pore pressure diffusion is rapid, and that the pore pressure can simply be modeled by $\alpha_f ^b$. Later, we will discuss the situation when erosion takes place for different dynamical variables, physical and material parameters and drift coefficients. Similarly, we show how physically relevant parameters and drift coefficients across the interface can avoid singularity in the erosion rate.

\subsubsection*{Reduced solid-erosion rates}
 
{\bf A.} If the solid and the fluid densities across the interface are similar $\lb \rho_s^b \approx \rho_s^m; \rho_f^b \approx \rho_f^m \rb$, then (\ref{MassProduction_Rate_d}) reduces to
\begin{eqnarray}
\begin{array}{lll}
\D{E_s =
\frac{\sqrt{\lb 1-\gamma\rb \lb\mu_s^m \alpha_s^m -  \mu_s^b\alpha_s^b \rb}}
{\sqrt{\nu\lb \lambda_{s_l}^m \alpha_s^m-  \lambda_{s}^b\alpha_s^b\rb}} \sqrt{g \cos\zeta h}
},
\end{array}    
\label{MassProduction_Rate_e}
\end{eqnarray}
where  $ 1-\gamma^m \approx 1-\gamma^b = 1-\gamma$.
\\[3mm]
{\bf B.} If in addition the friction coefficients are close to each other $\lb \mu_s^m \approx \mu_s^b \approx \mu \rb$ then, (\ref{MassProduction_Rate_d}) further reduces to:
\begin{eqnarray}
\begin{array}{lll}
\D{E_s =
\frac{\sqrt{\mu \lb 1-\gamma\rb\, \lb \alpha_s^m -  \alpha_s^b \rb}}
{\sqrt{\nu\lb \lambda_{s_l}^m \alpha_s^m-  \lambda_{s}^b\alpha_s^b\rb}} \sqrt{g \cos\zeta h}
}.
\end{array}      
\label{MassProduction_Rate_g}
\end{eqnarray}
Both cases take into account the solid fraction difference across the interface, which  evolves dynamically during the flow.

\subsection{Fluid erosion-rate}

Next, we develop the erosion rate for the fluid. Notations are analogously defined as for the solid erosion rate. For this, we again consider the basic relation between the stress and momentum flux jump for fluid across the singular surface induced by the erosion propagation normal to the erodible basal surface:   
\begin{eqnarray}
\begin{array}{lll}
\D{\tau_{f_l}^m -\tau_f^b = E_f\lb \rho_f^m u_{f_l}^m \alpha_f^m-  \rho_f^b u_{f}^b\alpha_f^b\rb}.
\end{array}    
\label{MassProduction_Rate_af}
\end{eqnarray}
Fluid shear stress is described by the Chezy-type relationship (Chow, 1973; Fraccarollo and Capart, 2002). So, the erosion rate can be expressed as
\begin{eqnarray}
\begin{array}{lll}
\D{E_f = \frac{\tau_{f_l}^m -\tau_f^b}{\lb \rho_f^m u_{f_l}^m \alpha_f^m-  \rho_f^b u_{f}^b\alpha_f^b\rb} 
= \frac{C_f^m\rho_f^m\alpha_f^m \lb u_{f}^m\rb^2 - \, C_f^b\rho_f^b \alpha_f^b\lb u_{f}^b\rb^2}
       {\lb \rho_f^m u_{f_l}^m \alpha_f^m-  \rho_f^b u_{f}^b\alpha_f^b\rb} 
= \frac{C_f^m\rho_f^m \alpha_f^m\lb u_{f}^m\rb^2 - \, C_f^b\rho_f^b \alpha_f^b\lb u_{f}^b\rb^2}
       {\lb \rho_f^m\lambda_{f_l}^m \alpha_f^m -  \rho_f^b \lambda_{f}^b\alpha_f^b\rb u_f}},
\end{array}    
\label{MassProduction_Rate_bf}
\end{eqnarray}
where $C_f^m, C_f^b$ are the Chezy friction coefficients. A shearing can develop in the fluid. To take this into account, (\ref{MassProduction_Rate_bf}) needs to be depth-averaged. This is achieved by: 
\begin{eqnarray}
\begin{array}{lll}
\D{E_f 
= \frac{\lb C_f^m\rho_f^m \alpha_f^m- C_f^b\rho_f^b \lambda_{f}^b\alpha_f^b\rb u_f^2}
       {{H}\lb \rho_f^m\lambda_{f_l}^m \alpha_f^m -  \rho_f^b \lambda_{f}^b\alpha_f^b\rb u_f} h
= \frac{\lb C_f^m\rho_f^m \alpha_f^m- C_f^b\rho_f^b\lambda_{f}^b\alpha_f^b\rb}
       {{H}\lb \rho_f^m\lambda_{f_l}^m \alpha_f^m -  \rho_f^b \lambda_{f}^b\alpha_f^b\rb } hu_f
},
\end{array}    
\label{MassProduction_Rate_cf}
\end{eqnarray}
where $E_f$ is now associated with the depth averaged velocity, and ${H}$ is a typical scale for the flow depth. So,
\begin{eqnarray}
\begin{array}{lll}
\D{E_f 
= \frac{\lb C_f^m\rho_f^m \alpha_f^m- C_f^b\rho_f^b\lambda_{f}^b\alpha_f^b\rb}
       {{H}\lb \rho_f^m\lambda_{f_l}^m \alpha_f^m -  \rho_f^b \lambda_{f}^b\alpha_f^b\rb } hu_f
= E_f^{\,ri} u_f 
},
\end{array}    
\label{MassProduction_Rate_df}
\end{eqnarray}
where
\begin{eqnarray}
\begin{array}{lll}
\D{E_f^{\,ri} 
= \frac{\lb C_f^m\rho_f^m \alpha_f^m- C_f^b\rho_f^b\lambda_{f}^b\alpha_f^b\rb h}
       {{H}\lb \rho_f^m\lambda_{f_l}^m \alpha_f^m -  \rho_f^b \lambda_{f}^b\alpha_f^b\rb},
}
\end{array}    
\label{MassProduction_Rate_ef}
\end{eqnarray}
is the erosion rate (intensity) factor for fluid. In general, (\ref{MassProduction_Rate_d}) and (\ref{MassProduction_Rate_df}) are the new mechanical models for the solid and fluid erosion rates, respectively.

\subsubsection*{Reduced fluid erosion-rate}

In simple situations, when a condition close to no-slip at the bed prevails, the fluid velocity at the bed could be negligible or, it can be much smaller than the mean fluid velocity. This means that $\lambda_f^b$ may be set identically equal to zero. In such a situation, (\ref{MassProduction_Rate_ef}) reduces to the simple expression for the erosion rate factor:
\begin{eqnarray}
\begin{array}{lll}
\D{E_f^{\,ri} 
= \frac{C_f^m}{{H}\lambda_{f_l}^m}\, h.
}
\end{array}    
\label{MassProduction_Rate_ff}
\end{eqnarray}
Other reduced but more general expressions than (\ref{MassProduction_Rate_ff}) can be obtained, as for the reduced solid erosion rates, by considering similarities in fluid densities, the friction coefficients, and/or the volume fractions across the interface.

\subsection{Further erosion rate models}

The solid and fluid erosion rate models developed so far are based on the explicit two-phase debris flow, and two-phase bed. These models are essentially based on the solid and fluid stresses, and solid and fluid momentum fluxes across the interface between the moving debris material and erodible bed. Further models for erosion rates can be readily obtained by combining stresses and momenta in different ways. For example, consider the two-phase debris dynamics with explicit shear stresses for the solid and fluid $\tau_{s_l}^m$ and $\tau_{f_l}^m$, but a bulk resisting shear stress for the basal erodible substrate, say $\tau_{B}^b$, representing both the solid and fluid stresses in the mixture. So, the jump in the shear stress is $\lb \tau_{s_l}^m + \tau_{f_l}^m\rb - \tau_{B}^b$. For the momentum fluxes, the above solid and fluid fluxes can be appropriately combined, either phase-wise (or, with barycentric velocities and densities). Then, however, instead of two different erosion rates, now, only one bulk erosion rate $E_B$ emerges. The solid and fluid erosion rates can still be expressed as $E_s = \alpha_s^b E_B$, and $E_f = \alpha_f^b E_B$. As it is straightforward, we do not further consider the full derivation here. Nevertheless, this involves two additional parameters, one representing the fluid pressure in the bed that appears due to Terzaghi effective stress, and the other parameter connecting fluid velocity to the solid velocity, say a velocity drift coefficient. However, both of these parameters might not have a simple mechanical constrain (Iverson and Denlinger, 2001; Pudasaini et al., 2005; Iverson, 2012). Further difficulties may appear in obtaining expressions for erosion drift coefficients $\lambda_{s_l}^m$ and $\lambda_{s}^b$, and the shear factor $\nu$. Moreover, the erosion rates thus constructed may, or may not be better than those obtained previously in Section 3.1 and Section 3.2. Further models can be developed that include cohesion. The models presented in Section 3.1 and Section 3.2 generalize and extend the bulk mixture erosion rate models in (Fraccarollo and Capart, 2002) and Iverson (2012) to two-phase erosion rate models.

\section{The erosion mechanics: stronger gains, weaker loses}

\subsection{Erosion enhances mobility: solving the dilemma of mass mobility}

Usually erosion related geophysical mass flows are more mobile than without erosion. However, this fact has never been explained mechanically explicitly and unambiguously. In literature, some mention that erosion results in shorter travel distance due to the energy lost in erosion (Le and Pitman, 2009; Crosta et al., 2013, 2015). It has also been argued that other than erosion there must be some further mechanisms causing higher mobility. While others present results showing that due to the added mass, the debris travels longer distance (Iverson, 1997; Hungr et al., 2005; McDougall and Hungr, 2005; Rickenmann, 2005; Godt and Coe, 2007; Mangeney et al., 2007, 2010; Bouchut et al., 2008; Reid et al., 2011). However, no clear explanation and derivation exists to mechanically explicitly describe the state of mobility. Incompatible and conflicting thoughts and results are presented (Chen et al., 2006;  Mangeney et al., 2007, 2010; Le and Pitman, 2009; Crosta et al., 2015). This is a long standing dilemma in mass flow mobility associated with erosion. We address this issue here.
\\[3mm]
The erosion resulting in higher geophysical mass flow mobility can be described mechanically, and quantitatively when the mass and momentum productions are consistently and physically correctly included into the mass and momentum equations. Consider the solid-type flow or, solid component in the mixture flow. The following descriptions also apply to fluid component, and two-phase mixture mass flows. We should always consider that erosion takes place only when the bed is mechanically weaker in relation to the flowing material itself. First, assume that no erosion takes place. Then, during the motion momentum is dissipated as frictional stress in the momentum balance by the Coulomb friction term, the second term on the right hand side of (\ref{Model_Final_ss}). Next, consider that due to the presence of a weaker bed erosion takes place. The relatively weaker bed introduces effectively reduced frictional stress. This reduced stress, which is a positive quantity, is balanced by the system consistently and rigorously, matching the additional (produced) momentum, and eroded mass. This appears systematically and exactly in the momentum balance (\ref{Model_Final}), the quantity $u_s^bE_s$ (the second term on the right hand slide of the first equation). Alternatively, mathematically this can also be explained as: mass production (erosion) leads to the corresponding momentum production, which is equivalent to the effectively reduced frictional stress. Yet, importantly, as the mass is added into the system, the gravity load immediately accelerates the total mass (initially triggered plus the newly added mass) down the entire travel distance. This further enhances the flow mobility, because the erosion-induced added mass implies added potential energy into the system. 
\\[3mm]
In Section 4.3 we show that $\tau_{s_l}^m - \tau_{s}^b$ is equivalent to $u_s^bE_s$. On the one hand, from the mechanical point of view, we could have reduced the shear stress due to erosion by replacing $\tau_{s_l}^m$ by $\tau_{s}^b$, where $\tau_{s_l}^m > \tau_{s}^b$, thus reducing the total stress by the amount $\tau_{s_l}^m - \tau_{s}^b$, and do not include the momentum production, $u_s^bE_s$. However, in this situation, only erosion is considered, and not the entrainment and transport of the eroded mass. On the other hand, the incorporation of the kinematic and dynamic boundary conditions indicates that inclusion of the mass and momentum productions into the mass and momentum balance equations is essential, and while doing this, we do not need to additionally adjust the shear stress in erosion, the shear stress $\tau_{s_l}^m$ works compatibly with the momentum production, $u_s^bE_s$. The basic models (\ref{Model_Final_Mass})-(\ref{Model_Final_fy}) are developed consistently following these mechanical processes. In total, the momentum production $u_s^b E_s$, and the gained potential energy due to added mass that induces extra driving gravity load $\rho_s^b\alpha_s^b\, h_E \, g \sin\zeta$, where $h_E$ is the erosion depth (or area, or volume, chosen consistently with dimension of the problem) results in higher mobility. The actual total gravity load is  $\D{\alpha_s^m h \, g \sin\zeta\lb 1 +  \frac{\rho_s^b\alpha_s^b h_E}{\rho_s^m\alpha_s^m h}\rb}$. This explains the higher mobility associated with the erosion in mass flow. So, we solved the long standing dilemma of erosion related mobility of geophysical mass flows. With respect to the arguments of erosion induced decreased mass flow mobility, this is (or, can be perceived as) a paradigm shift.

\subsection{Deposition as reverse erosion process}

Next, we show that mechanically deposition is the reversed process of erosion. Another unsolved problem in mass flow is about deposition. Some mention that deposition can be modelled just by considering the negative of the erosion process (McDougall and Hungr, 2005; Iverson, 2012). Others dispute that deposition process cannot always be described this way (see, e.g., Issler, 2014). However, there exists no clear derivation and explanation for why this is mechanically right, or not right, to consider deposition as negative process of erosion.  One may think of the rocketing effect when the mass is lost in deposition if this is just described as the reverse process of erosion. However, this is not the situation in mass flow as the deposition results from the external shear resistance from the basal material (Hungr, 1990; Iverson, 2012). Here, we make it clear, that fundamentally the same process can also be applied to physically correctly describe the deposition which is achieved by consistently reversing the erosion process. The deposition process begins not when the sliding mass starts to lose (leave) some portion of its mass and give it to the basal surface that, as we may think, would lead to rocketing (thrust) resulting in acceleration (which is the wrong concept in mass flow). But, the deposition begins as soon as the bed starts to decelerate (the frontal part or, the lower layer of) the flow due to the higher frictional resistance of the bed than that of the flowing material. Deposition requires further conditions to be fulfilled. Dynamically deposition follows decelerating state, i.e., ${\partial u}/{\partial t} < 0$. Then, deposition preferentially takes place, e.g., where the bed friction angle is higher than the slope angle, i.e., $\delta>\zeta$. Thus, during deposition, as the basal surface starts to gain the mass, it results in relatively increased shear stress in the deposition area, $\tau_s^b$, which is larger than $\tau_{s_l}^m$. This effectively increased frictional stress, $-(\tau_{s_l}^m - \tau_{s}^b)$, is mechanically and appropriately balanced by the momentum loss, $-u_s^b E_s$, in the momentum balance equation. Here, the negative sign appears due to the reverse process $-E_s$, as also clearly seen in $-(\tau_{s_l}^m - \tau_{s}^b)$ for deposition as compared to $(\tau_{s_l}^m - \tau_{s}^b)$ for erosion. Fundamentally the same mechanical process of erosion applies, but reversely, to the deposition process. So, `stronger gains, weaker loses' $-$ this is the mechanical process in geophysical mass flow. That is, during the erosion mass is gained by the flow, whilst during deposition mass is gained by the basal substrate. 
\\[3mm]
As the mass enters into the run-out zone, or any mechanically stronger region, that takes into account the above mentioned further conditions, deposition process is triggered, may be weakly, slowly, or rapidly. But, then due to deposition process, the deposited material is compacted, interlocking among the grain develops, effective friction increases as the mass transforms from dynamic to static state (Pudasaini et al., 2007) and as the relatively large amount of fluid (if exists in the mixture) runs off, and the solid volume fraction may increase substantially. Furthermore, the driving gravity force is reduced (in transition), or largely reduced (in run-out zone). All these result in enhancing the effective strength of the material in the dynamically evolving basal surface. Then, when some or all of these mechanisms are in effect, the deposition process amplifies. The deposition process will reach its final stage, and the mass fully comes to a standstill as soon as $h_s \to 0$.  

\subsection{Balance of effective reduced frictional stress and momentum production in erosion}

Now, we prove that the effectively reduced frictional dissipation and momentum production are equivalent. First, assume the non-erosional situation for which the shear resistance against the motion of the debris is $-\tau_{s_l}^m = -\lb 1-\gamma^m\rb\rho_s^m g\cos\zeta h \mu_s^m \alpha_s^m$. Next, assume the situation when erosion takes place. The basal substrate applies the shear stress $-\tau_s^b = -\lb 1-\gamma^b\rb\rho_s^b g \cos\zeta h \mu_s^b \alpha_s^b$ on the sliding debris. Therefore, the momentum balance must be adjusted by replacing $-\tau_{s_l}^m$ by $-\tau_{s}^b$ in the corresponding Coulomb frictional resistance. And, the difference $-\tau_{s}^b - (-\tau_{s_l}^m) = \tau_{s_l}^m - \tau_{s}^b$, as compared to $\tau_{s_l}^m$, constitutes the net-momentum gain. The reduction in the Coulomb frictional dissipation (i.e., momentum gain) is $\tau_{s_l}^m - \tau_{s}^b = g\cos\zeta h\left[\lb 1-\gamma^m\rb\rho_s^m\mu_s^m\alpha_s^m - \lb 1-\gamma^b\rb\rho_s^b\mu_s^b\alpha_s^b\right]/\rho_s^m\alpha_s^m$, where $\rho_s^m\alpha_s^m$ in the denominator appears due to the mass fraction factor in the momentum balance equation. Whereas, from (\ref{MassProduction_Rate_c}) the momentum production is $u_s^bE_s = \lambda_s^b g\cos\zeta h\left[\lb 1-\gamma^m\rb\rho_s^m\mu_s^m\alpha_s^m - \lb 1-\gamma^b\rb\rho_s^b\mu_s^b\alpha_s^b\right]/(\rho_s^m\lambda_{s_l}^m\alpha_s^m - \rho_s^b\lambda_s^b\alpha_s^b)$, where the factor $\lambda_s^b$ in the numerator appears due to the term $u_s^b/u_s$ in the momentum production. These two expressions are connected in two ways. Either we can consider that these two momentum exchange quantities are equal (or, equivalent), or for a particular choice of $\lambda_s^b$ these two expressions are equal. In both cases, we obtain the erosion drift (coefficient) equation: 
\begin{eqnarray}
\lambda_{s_l}^m = \lb 1+ \D{\frac{\rho_s^b}{\rho_s^m}\frac{\alpha_s^b}{\alpha_s^m}}\rb \lambda_{s}^b.
\label{lambda_m}
\end{eqnarray}
This is a unique expression connecting $\lambda_{s}^b$ with $\lambda_{s_l}^m$ via the ratios of the densities ${\rho_s^b}/{\rho_s^m}$, and the volume fractions $\alpha_s^b/\alpha_s^m$ on either side of the interface. Since both $\lambda_{s_l}^m$ and $\lambda_{s}^b$ are bounded from above by unity, for erosional configuration, $\alpha_s^m$ must be suitably bounded away from zero from below (see later).
\\[3mm]
This clearly proves (or, demonstrates) that the erosional configuration reduces the frictional dissipation. This can be included into the system either by appropriately replacing $-\tau_{s_l}^m$ by $-\tau_{s}^b$ in the frictional dissipation term, or simply through momentum production term that emerges formally and systematically. The momentum gained by effectively reducing the frictional stress $(\tau_{s_l}^m -\tau_{s}^b)$ is equivalent to the momentum production $u_s^b E_s$. So, without considering the momentum gain by reducing the friction, i.e., by simply using the Coulomb friction $-\tau_{s}^b - (\tau_{s_l}^m -\tau_{s}^b) = -\tau_{s_l}^m$, the momentum production $u_s^b E_s$ introduces additional momentum into the system. This means, non-erosional setting loses more momentum in friction than the erosional setting. So, relatively, momentum is gained in erosion. This effective momentum production results in higher mobility.
\\[3mm]
Similar analysis also applies for fluid, for which the frictional resistance is quadratically proportional to the flow velocity (ambient viscous drag). Without involving these complications of finding suitable pre-existing terms in the momentum balance equations, i.e., the Coulomb and viscous drag, by simply following the rigorous derivations, the gained momentum, due to effectively reduced friction in erosional setting, can be systematically included into the system via the momentum production terms for both the solid and fluid phases.
\\[3mm]
 The erosion drift coefficient for the  fluid phase can be derived similarly to equation (\ref{lambda_m}). In simple situation when $h/H \approx 1$, we obtain: 
\begin{eqnarray}
\lambda_{f_l}^m
= \lb \frac{h}{H}+ \D{\frac{\rho_f^b}{\rho_f^m}\frac{\alpha_f^b}{\alpha_f^m}}\rb \lambda_{f}^b
\approx \lb 1+ \D{\frac{\rho_f^b}{\rho_f^m}\frac{\alpha_f^b}{\alpha_f^m}}\rb \lambda_{f}^b. 
\label{lambda_m_f}
\end{eqnarray}
 As for the solid, for erosional configuration, $\alpha_f^m$ must be suitably bounded away from zero from below.

\section{Discussions on erosion drifts, shear velocity factor and erosion-rates}

\subsection{Erosion drifts}

The two different erosion drift coefficients $\lambda_{s_l}^m$ and $\lambda_{s}^b$ emerge due to the jump in velocity on either side of the mobile singular surface. For erosion to take place, these parameters must appear, in general be distinct, or non-zero and positive. This is a result of the triggering requirement for the erosion, i.e., the applied shear stresses on either side of the singular surface are not equal, and that there is a jump in the momentum flux. For mechanically strong bed material $\lambda_{s}^b$ is very small. The value of $\lambda_{s_l}^m$, however, depends on the velocity profile of the moving material. For strong basal shearing $\lambda_{s_l}^m$ is smaller, but for more plug-like flow, $\lambda_{s_l}^m$ is close to unity. For weaker bed material $\lambda_{s}^b$ may be large, but could still be much smaller than $\lambda_{s_l}^m$. This can be better explained with the erosion drift equation (\ref{lambda_m}) with several very important implications. 
\\[3mm]
{\bf I.} The densities $\lb \rho_s^b, \rho_s^m\rb$ and the volume fractions $\lb \alpha_s^b, \alpha_s^m\rb$ are positive quantities. Since the particles in the debris motion slip along the basal surface, the velocity of the eroded particle must be non-zero positive and this velocity must be smaller than the velocity of the flow. Thus, the erosion drift equation (\ref{lambda_m}) implies that $\lambda_{s_l}^m > \lambda_{s}^b$, and $\lambda_{s}^b > 0$. So, it is mechanically incorrect to set $\lambda_{s}^b = 0$, or $u_s^b = 0$. This is intuitively clear and natural condition for erosion to take place, but contradicts prevailing considerations (Fraccarollo and Capart, 2002; Iverson, 2012).
\\[3mm]
{\bf II.} Rapid erosion (as $\lambda_s^b \to \lambda_{s_l}^m$ or, $\lambda_s^b \to 1$) is a result of a mechanically significantly weaker basal material than the flowing material (in terms of the density, and/or the bed that consists of a fluid-type material, i.e., small $\alpha_s^b$).
\\[3mm]
{\bf III.} If ${\rho_s^b} < {\rho_s^m}$ and ${\alpha_s^b} < {\alpha_s^m}$, or ${\rho_s^b\alpha_s^b} < {\rho_s^m\alpha_s^m}$ then, $\lambda_{s}^b > \D{\frac{1}{2}}\lambda_{s_l}^m$. This situation appears when the bed is mechanically weaker than the flowing material (in terms of the solid densities and volume fractions, or the mass fraction component) and implies that the eroded particle moves with relatively higher velocity. On the other hand, if ${\rho_s^b} > {\rho_s^m}$ and ${\alpha_s^b} > {\alpha_s^m}$, or ${\rho_s^b\alpha_s^b} > {\rho_s^m\alpha_s^m}$ then, $\lambda_{s}^b < \D{\frac{1}{2}}\lambda_{s_l}^m$. So, when the bed is stronger as compared to the flow, in terms of the mass fraction component, then the eroded particle moves with relatively slower velocity. 
 These are intuitively clear phenomena and imply the most simple possible numerical domains for $\lambda_{s_l}^m$ and $\lambda_{s}^b$: $\lambda_{s_l}^m \in (0, 1)$ and $\lambda_{s}^b\in (0, 1)$ or, $\lambda_{s}^b\in (0, 1/2)$.
\\[3mm]
{\bf IV.} In simple situations for which the solid densities and volume fractions on either side of the interface are the same (i.e., ${\rho_s^b}= {\rho_s^m}; {\alpha_s^b} = {\alpha_s^m}$), then (\ref{lambda_m}) reduces to $\lambda_{s}^b = \D{\frac{1}{2}}\lambda_{s_l}^m$. This means the velocity of the eroded particle is one half of the velocity of the particle at the base of the debris flow.
\\[3mm]
{\bf V.} The coefficient $\lambda_{s_l}^m$ provides information on the velocity profile through the flow depth that, in applications, can be assumed to be parabolic or plug-type. Plug flow consideration is compatible with the depth-averaged shallow flows. For the simple plug-type flow $\lambda_{s_l}^m \approx 1$, which, for $\rho_s^b = \rho_s^m; {\alpha_s^b} = {\alpha_s^m}$, implies $\lambda_s^b = 1/2$. Such a value has also been obtained by Le and Pitman (2009) by utilizing kinetic theory (Jenkins and Savage, 1983). So, the drift equation (\ref{lambda_m}) appears to be rich in explaining the erosion dynamics. Thus, with the knowledge of the velocity profile, or the plug flow, (\ref{lambda_m}) is closed because, $\alpha_s^m$ is a dynamical quantity, $\alpha_s^b$ is known or, another dynamical quantity, and the densities are known quantities. 
\\[3mm]
{\bf VI.} Similar analysis holds for the fluid. In general, both $\lambda_{f_l}^m$ and $\lambda_{f}^b$ can be (much) smaller than  $\lambda_{s_l}^m$ and $\lambda_{s}^b$, respectively. Since the fluid constituent in the flowing debris may consist of a mixture of water, silt, clay and other fine particles, in general $\rho_{f}^m > \rho_{f}^b$, as the fluid in the bed material is mostly pure water. With $\alpha_f^m > \alpha_f^b$, this implies that $\lambda_{f}^b > \D{\frac{1}{2}}\lambda_{f_l}^m$. Nevertheless, $\lambda_{f}^b \approx \D{\frac{1}{2}}\lambda_{f_l}^m$ holds when the flowing debris fluid is mostly water.

\subsection{Shear velocity factor}

The shear velocity factor $\tilde\nu$ is the transformation factor between the effective (relative) erosional shear stress and effective velocity jump in erosional situation. Depending on the density jump, it can be large when $\lambda_{s_l}^m - \lambda_{s}^b$ is small. In simple situations, some typical values of $\tilde\nu$ could be considered. Next, we derive an expression for $\nu$. For this, consider the balance between the momentum gained by effectively reduced Coulomb friction, and the momentum production $u_s^b E_s$ from (\ref{MassProduction_Rate_d}):
\begin{eqnarray}
\D{\frac{1}{\rho_s^m\alpha_s^m}}\bigg[\! \lb 1\!-\! \gamma^m\rb \!\rho_s^m \mu_s^m \alpha_s^m \!-\! \lb \!1\!-\! \gamma^b\rb\!\rho_s^b \mu_s^b\alpha_s^b \bigg] g \cos\zeta h 
\!=\! \frac{\sqrt{ \!\lb 1\!-\! \gamma^m\!\rb\rho_s^m \mu_s^m \alpha_s^m \!-\! \lb 1\!-\! \gamma^b\!\rb\rho_s^b \mu_s^b\alpha_s^b \,}}
{\sqrt{\nu \lb\rho_s^m \lambda_{s_l}^m\alpha_s^m \!-\!  \rho_s^b \lambda_{s}^b\alpha_s^b\,\rb}} \sqrt{g \cos\zeta h}\, \lambda_s^b u_s.
\label{Factor_nu}
\end{eqnarray}
As we are interested in the order of magnitude estimation for the factor $\nu$, we apply the ordering for $h$ and $u_s$, i.e, typical flow depth $H$ and the typical flow velocity $\sqrt{gL}$, and consider the relevant ordering. With $\varepsilon = H/L$, (\ref{Factor_nu}) reduces to
\begin{eqnarray}
\D{\frac{1}{\nu}} \approx \D{\frac{\varepsilon\cos\zeta\left[ \lb 1- \gamma^m\rb\rho_s^m \mu_s^m - \lb 1- \gamma^b\rb\rho_s^b \mu_s^b\,\alpha_s^b/\alpha_s^m\right] \lb \rho_s^m \lambda_{s_l}^m -  \rho_s^b \lambda_{s}^b\,\alpha_s^b/\alpha_s^m\rb }{\lb \rho_s^m\lambda_{s}^b\rb^2}}.
\label{Factor_nu_a}
\end{eqnarray}
In simple situations when the densities and the friction coefficients across the interface are similar (i.e., $\rho_s^m \approx \rho_s^b, \rho_f^m \approx \rho_f^b;\, \mu_s^m \approx \mu_s^b = \mu$), and if the plug-type flow is considered ($\lambda_{s_l}^m \approx 1$), (\ref{Factor_nu_a}), with (\ref{lambda_m}), reduces to
\begin{eqnarray}
\D{\frac{1}{\nu}} = \varepsilon (1-\gamma)\cos\zeta \mu \lb 1 - \lb{\alpha_s^b}/{\alpha_s^m}\rb^2 \rb,
\label{Factor_nu_b}
\end{eqnarray}
which is positive for $\alpha_s^m > \alpha_s^b$. If $\lambda_{s}^b$ takes the value close to 1/2, then (\ref{Factor_nu_a}) can be further reduced to have some idea about the order of magnitude estimate for $\nu$: $1/\nu \approx 2\varepsilon (1-\gamma)\cos\zeta \mu \lb 1 - {\alpha_s^b}/{\alpha_s^m} \rb$. These show that, if in addition, the volume fractions (concentrations) across the interface are similar, erosion does not take place. For the considered simplifications erosional state automatically requires that $\alpha_s^m > \alpha_s^b$. So, as $\alpha_s^b > 0$ in realistic situations, no singularity appears.

\subsection{Erosion-rates}

Here, we discuss some important physical and mechanical aspects of the solid erosion model (\ref{MassProduction_Rate_d}). Together with the closure for the shear velocity factor (\ref{Factor_nu_a}) the solid erosion rate reduces to:
\begin{eqnarray}
\begin{array}{lll}
\D{E_s =
\frac{ \lb 1- \gamma^m\rb\rho_s^m \mu_s^m - \lb 1- \gamma^b\rb\rho_s^b \mu_s^b\,\alpha_s^b/\alpha_s^m}
{\rho_s^m\lambda_{s}^b} \, \cos\zeta \sqrt{{\varepsilon\,gh}}},
\end{array}    
\label{E_s_final}
\end{eqnarray}
The model (\ref{MassProduction_Rate_d}) or, (\ref{E_s_final}) explicitly shows, that for erosion to take place, at least one of the four parameters or, the dynamical variables, must have a jump across the interface. These are, the buoyancy factors $\lb 1- \gamma^m, 1- \gamma^b\rb$, the solid densities ($\rho_s^m, \rho_s^b$), friction coefficients ($\mu_s^m, \mu_s^b$), and the volume fractions ($\alpha_s^m, \alpha_s^b$). The two-phase mass flow consideration made it possible to include these effects because all these variables and parameters differ between solid and fluid phases resulting in fundamentally different erosion rates for solid and fluid with corresponding $1-\gamma, \rho, \mu$, and $\alpha$ values. 
\\[3mm]
For bed material with a similar density to the flowing material ($\rho_s^m = \rho_s^b, \rho_f^m = \rho_f^b$) and an erosion drift coefficient of $\lambda_{s_l}^m = 1$, the erosion rate (\ref{E_s_final}), with (\ref{lambda_m}), reduces to:
\begin{eqnarray}
\begin{array}{lll}
\D{E_s = \lb 1-\gamma\rb 
\lb \mu_s^m  -  \mu_s^b\,\frac{\alpha_s^b}{\alpha_s^m} \rb \cos\zeta \sqrt{{\varepsilon\,gh}}\lb 1 + \frac{\alpha_s^b}{\alpha_s^m}\rb }.
\end{array}    
\label{E_s_final_red}
\end{eqnarray}
Further reduction is possible if the frictional (or, mechanical) similarity prevails ($\mu_s^m = \mu_s^b = \mu$),
\begin{eqnarray}
\begin{array}{lll}
\D{E_s = \lb 1-\gamma\rb \mu \lb 1 - \lb{\alpha_s^b}/{\alpha_s^m}\rb^2 \rb \cos\zeta \sqrt{{\varepsilon gh}}
= \left [ \lb 1-\gamma\rb \mu \lb 1 - \lb{\alpha_s^b}/{\alpha_s^m}\rb^2 \rb \cos\zeta \sqrt{\varepsilon g}\right ]
 \sqrt{h}
= E_{s_e}^{ri}\sqrt{h}
}.
\end{array}    
\label{E_s_final_maxred}
\end{eqnarray}
 If $\lambda_{s}^b \approx 1/2$ is appropriate, then $E_{s_e}^{ri}$ further reduces to: $E_{s_e}^{ri} = 2\lb 1-\gamma\rb \mu\lb 1 -{\alpha_s^b}/{\alpha_s^m} \rb\cos\zeta \sqrt{\varepsilon g}$. This shows that the effective erosion rate intensity factor $E_{s_e}^{ri}$ is essentially driven by the contrast in the volumetric concentrations of the solid in the flowing material and the bed. $E_{s_e}^{ri}$ attains bounded values even for very low solid concentration, say $\alpha_s^m = 0.05$. For the typical values of $\rho_{f} =  1,100$ kg m$^{-3}$,\, $\rho_ s = 2,700$\, kg m$^{-3}$, $\delta = 25^\circ, \zeta = 45^\circ, \varepsilon = 3.0 \times 10^{-3}$ (see Section 6.1 for parameter choice), $E_{s_e}^{ri}$ is on the order of $10^{-2} - 10^{-1}$. Typically, $E_{s_e}^{ri} \approx 0.003$ for a reasonable choice of $\alpha_s^m = 0.63$ and $\alpha_s^b = 0.60$. If the solid concentration in the mixture is even below $\alpha_s^m = 0.05$ then, its effect in the mixture can be effectively neglected, because, in such a particle-laden, very dilute flow, due to the relatively large mean free paths for the solid particles, the grain frictional effects remain insignificant and the flow behaves as macroviscous (Pudasaini, 2011). So, essentially, even the reduced model (\ref{E_s_final_maxred}) does not contain singularity. However, in general, the basic model (\ref{MassProduction_Rate_d}) should be considered that avoids singularity due to the complex compositions of material parameters and dynamical variables. 
\\[3mm]
Due to the presence of different magnitudes in the physical parameters and the dynamical variables across the interface (i.e., the solid and fluid densities and volume fractions) erosion may occur even for the seemingly not possible but plausible relation $\mu_s^b > \mu_m^b$ for friction coefficients. Because usually one thinks only frictionally weaker basal layer could be eroded. The further important aspects of the new model (\ref{MassProduction_Rate_d}) are as follows. 
\\[3mm]
{\bf A.} It appears that erosion is not necessarily and directly dependent on the velocity but depends on the competition between the mechanical strength of the flowing debris and bed material: $\lb\!\lb \!1 \!-\! \gamma^m\!\rb\!\rho_s^m \mu_s^m \alpha_s^m \!-\! \lb \!1 \!-\! \gamma^b\!\rb\!\rho_s^b \mu_s^b\alpha_s^b\rb$. So, the erosion can take place even in a relatively slow movement, and may not necessarily take place even in rapid mass movements. Examples include, a strong bed with very low pore fluid pressure, or totally unsaturated, dry substrates. The erosion process is primarily governed by the material properties of the moving debris and bed, and the dynamical variables.  
\\[2mm]
{\bf B.} The erosion magnitude is proportional to the material load, $g \cos\zeta h$, and inversely proportional to the effective net density $\lb \rho_s^m \lambda_{s_l}^m \alpha_s^m -  \rho_s^b \lambda_{s}^b\alpha_s^b\rb$, and the typical velocity factor $\nu$ close to the bed.  Analytical closers for the coefficients $\lambda_{s_l}^{m}$, $\lambda_s^b$ and $\nu$, have been derived in Section 4.3 and Section 5.2. 
\\[2mm]
{\bf C.} $\sqrt{g\cos\zeta\, h\,}$ in (\ref{MassProduction_Rate_d}) has exactly the dimension of velocity. So, no bold and odd dimension appears as in the empirical models: $E_{emp}^{ri}\, h u$, in which $E_{emp}^{ri}$ has a dimension of [1/m].
\\[2mm]
{\bf D.} Erosion initiates mechanically if {$\left[\lb 1 \!-\!\gamma^m\rb\rho_s^m \mu_s^m \alpha_s^m \!-\! \lb 1 \!-\!\gamma^b\rb\rho_s^b \mu_s^b\alpha_s^b\right]\!>\! 0$}. Then, the erosion velocity, $\sqrt{g \cos\zeta h}$ is amplified by the overall factor {${\sqrt{ \lb 1 \!-\!\gamma^m\rb\rho_s^m \mu_s^m \alpha_s^m \!-\! \lb 1\!-\! \gamma^b\rb\rho_s^b \mu_s^b\alpha_s^b\,}} /{\sqrt{\nu( \rho_s^m \lambda_{s_l}^m \alpha_s^m\!-\!  \rho_s^b \lambda_{s}^b\alpha_s^b)}}$} that depends on $\alpha_s^m$, and $\alpha_s^b$ which evolve with the dynamics of the fluid in the debris material and erodible bed.
\\[3mm]
Similar discussions can be obtained for the fluid erosion-rate (\ref{MassProduction_Rate_df}). However, since the fluid flow dynamics and the mechanical responses are different from the solid, the fluid erosion rate is also fundamentally different. This is expressed in (\ref{MassProduction_Rate_df}) which shows that the fluid erosion-rate (erosion velocity) is proportional to the mean fluid velocity of the moving debris mass. Whereas the erosion rate (intensity) factor $E_f^{ri}$ includes all the fluid dynamical and mechanical variables and the physical parameters. These are the flow height, and the Chezy parameters, fluid densities, volume fractions, and the erosion drifts associated with the fluid. Erosion takes place as long as $\lb C_f^m\rho_f^m \alpha_f^m- C_f^b\rho_f^b\alpha_f^b\lambda_{f}^b\rb > 0$. As for the solid, the typical values of the erosion-drifts can be estimated. The Chezy parameters are determined by the flow configuration, or can also be chosen as some suitable numerical parameters (Chow, 1973; Fraccarollo and Capart, 2002). $C_f^m \approx 0.004$ is a reasonable value. Further $H \approx 2.0$ is an admissible choice. Then, for the erosional setting, typical value of the fluid related parameter can be chosen as $\lambda_{f_l}^m = 0.9$. With this, from (\ref{MassProduction_Rate_ff}) the effective fluid erosion rate intensity factor can be estimated as $E_{f_e}^{ri} = 0.002$, where the flow height $h$ is included in the dynamic simulation.

\section{Application of the new model and simulation results}

\subsection{Numerical method, simulation set-up and parameters}

The model equations (\ref{Model_Final_Mass})-(\ref{Model_Final_fy}) are a set of well-structured, non-linear hyperbolic-parabolic partial differential equations in conservative form with complex source terms. These model equations are used to compute the total depth $h$, solid volume fraction $\alpha_{s}$, velocity components for solid $\left(u_{s}, v_{s}\right)$ and fluid $\left(u_{f}, v_{f}\right)$ in $x$- and $y$-directions, respectively, and the evolution of the erodible (depositional) basal surface, $b$, as functions of space and time. The model equations are solved in conservative variables $\textbf{W}$ = $( h_{s}, h_{f} , b, m_{x_s}, m_{y_s}, m_{x_f}, m_{y_f})^{t}$, where $ h_{s}=\alpha_{s}h $,  $ h_{f}=\alpha_{f}h$ are the solid and fluid contributions to the debris mixture, or the flow height; and $(m_{x_s}, m_{y_s})  =  (\alpha_s h u_{s}, \alpha_s h v_{s}$), $(m_{x_f}, m_{y_f})  = (\alpha_f h u_{f}, \alpha_f h v_{f})$, are the solid and fluid momenta. This facilitates numerical integration even when shocks are formed in the field variables (Pudasaini, 2012; Kattel et al., 2016). {Shock formation is an essential mechanism in geophysical mass flows when the flow becomes subcritical from its supercritical state (Pudasaini, 2014). It is therefore natural to employ conservative high-resolution numerical techniques that are able to resolve the steep gradients and moving fronts often observed in experiments and field events but not captured by traditional finite difference schemes. So, simulations are performed with a high-resolution shock-capturing nonoscillatory central differencing, total variation diminishing scheme (Nessyahu and Tadmor, 1990; Tai et al., 2002; Pudasaini and Hutter, 2007).} Advantages of the applied unified simulation technique and the corresponding computational strategy have been explained in Pudasaini (2014), Kafle et al. (2016) and Kattel et al. (2016) for the two-phase subaerial debris flows, glacial lake outburst floods, submarine flows and subsequent tsunamis. For a better interpretation the simulations are performed in dimensional form. As simulation domain, we consider a two-dimensional debris flow down an inclined channel. The initial uniformly distributed, homogeneous mixture debris is released from the top of the channel. The parameter values chosen for simulation are: $\zeta = 45^\circ$, $\phi= 35^\circ$, $\delta = 25^\circ$,  $\rho_{f} =  1,100$ kg m$^{-3}$,\, $\rho_ s = 2,700$\, kg m$^{-3}$,\, $N_{R} = 30,000$,\, $N_{R_\mathcal{A}} = 1,000$,\, $Re_{p} = 1$,\, $\mathcal{U}_{T} = 1.0$ m s$^{-1}$,\, $\mathcal{P} = 0.75$,\, $\jmath = 1$,\, $\chi = 3$,\,  $\xi = 5$,\, $\mathcal{C} = 0.5, C_{VD} = 0.02$. {These parameter selections are based on the physics of the two-phase mass flows (Pudasaini, 2012, 2014; Pudasaini and Krautblatter, 2014; Kafle et al., 2016; Kattel et al., 2016).} As estimated in Section 5.3, the effective erosion rate intensity factors for the solid and fluid are (0.003, 0.002). These factors are obtained with some appropriate combinations of physically meaningful values of $\rho, \mu, \alpha, \nu, \lambda, \zeta$, etc., for solid, and corresponding physical parameters for fluid.

\subsection{Erosion and frontal surge dynamics, and effects of mass and momentum productions}

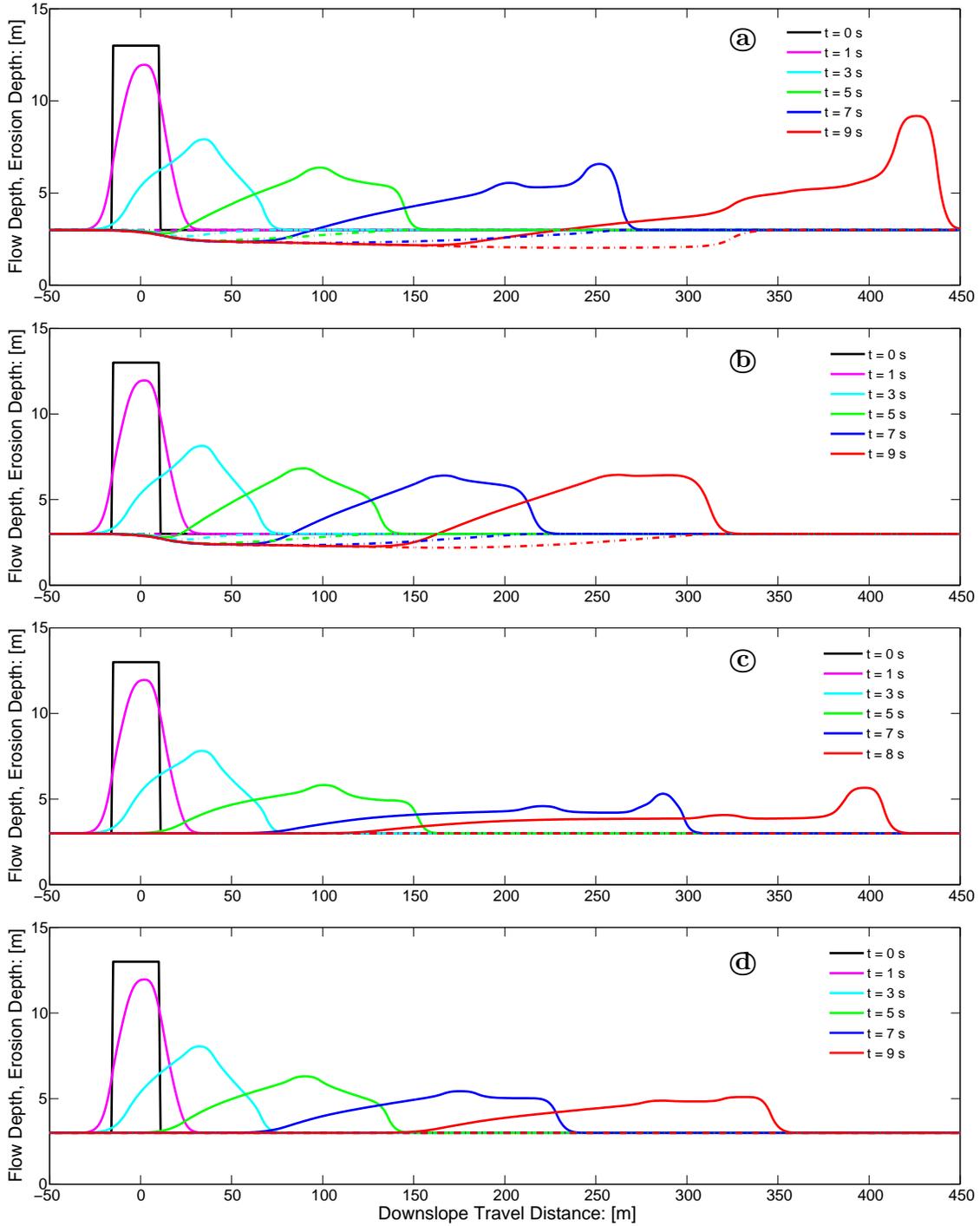
\begin{figure}[h!]
\begin{center}
\includegraphics[angle=0,width=18cm]{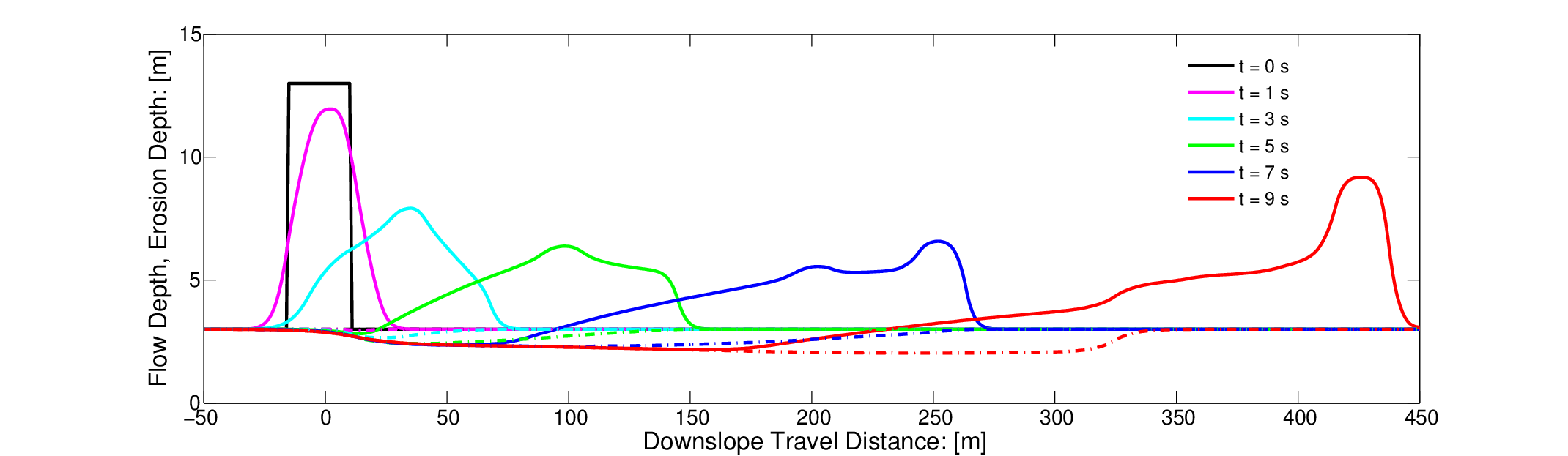}\\[-3.5mm]
\includegraphics[angle=0,width=18cm]{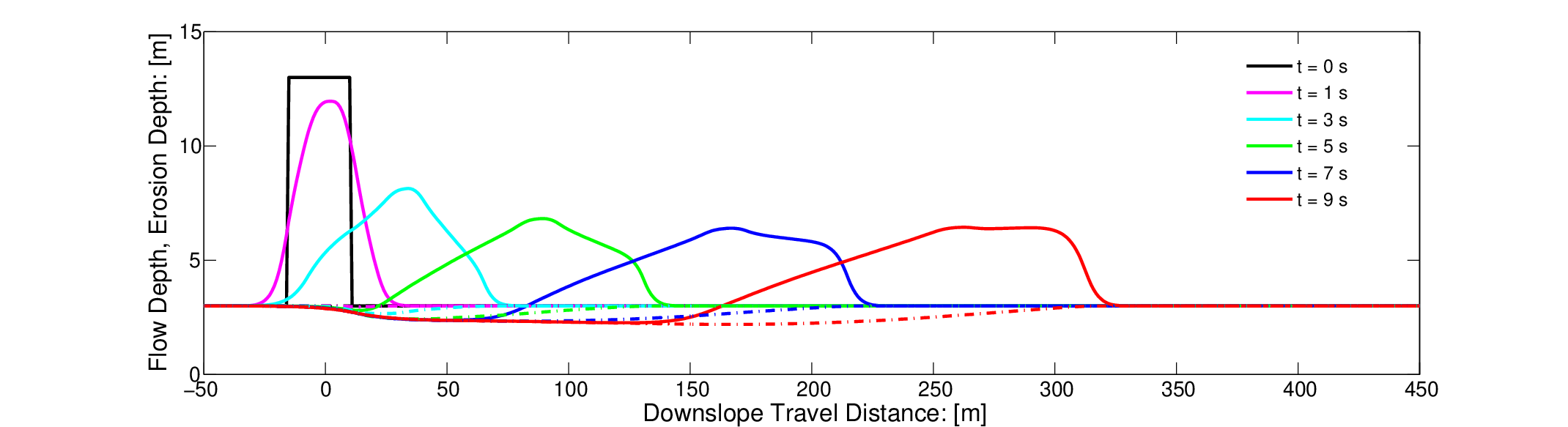}\\[-3.5mm]
\includegraphics[angle=0,width=18cm]{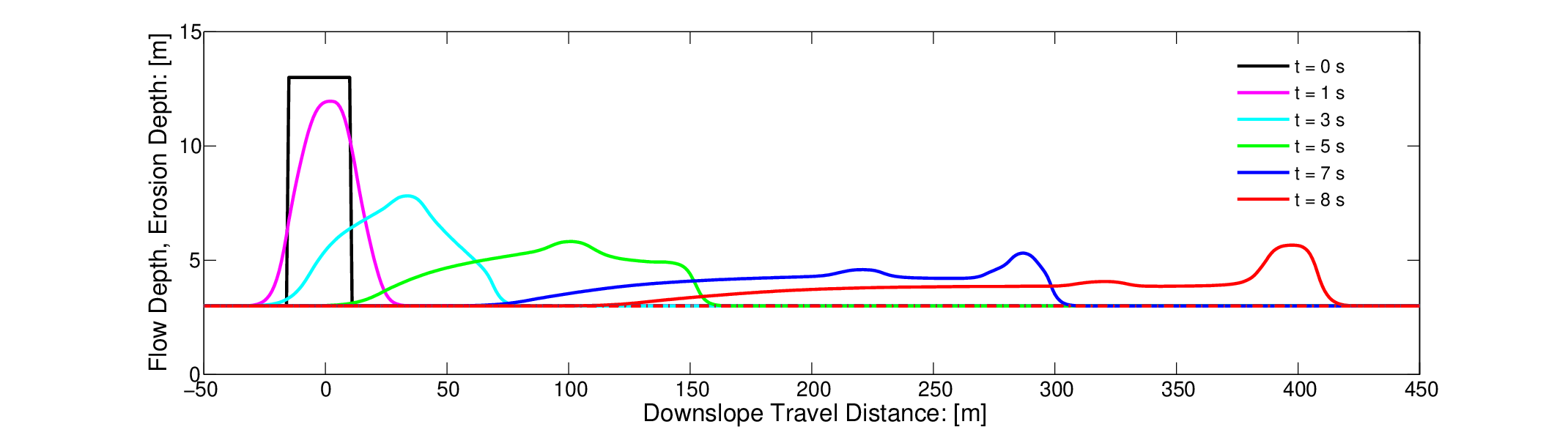}\\[-3.5mm]
\includegraphics[angle=0,width=18cm]{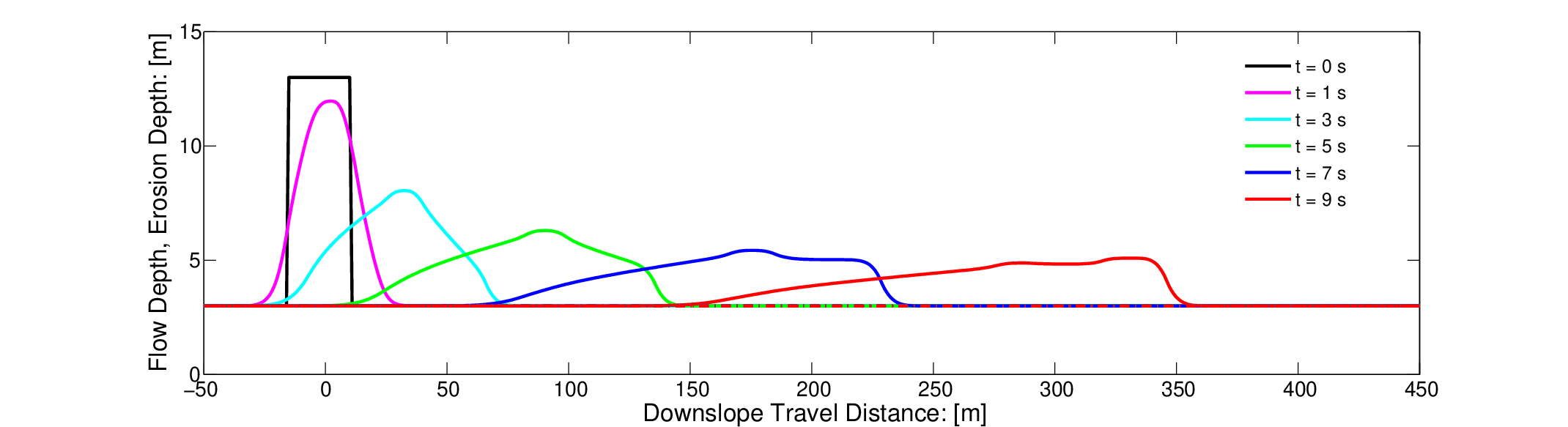}
\end{center}
\caption{Time evolution of the flow depth (solid line) and erosion depth (dashed-dot line) as a two-phase debris mass moves down an inclined erodible slope. Panels represent: {\bf \textcircled{a}} erosion with mass and momentum production, {\bf \textcircled{b}} physically incomplete erosion disregarding momentum production, {\bf \textcircled{c}} no erosion but momentum production (which is equivalent to an effectively reduced friction), and {\bf \textcircled{d}} no erosion, respectively. This reveals that momentum production results in higher flow mobility. Panels {\bf \textcircled{b}} and {\bf \textcircled{c}} are physically irrelevant.}
\label{Fig_3}
\setlength{\unitlength}{1.cm}
\begin{picture}(0,0)
\put(13,21.7){\bf  \textcircled{a}}
\put(13,16.77){\bf  \textcircled{b}}
\put(13,12.17){\bf \textcircled{c}}
\put(13,7.53){\bf  \textcircled{d}}
\end{picture}
\end{figure}
Figure \ref{Fig_3} shows the time evolution of the flow depth and erosion depth  as a two-phase debris mass moves down an inclined erodible slope. A 3 m erodible layer extends from $x = 10$ m to 325\,m along the downslope travel distance, consisting of 35\% pore space filled with fluid. Initial debris material ($t = 0$, rectangle) consists of 65\% solid and 35\% fluid. Evolving bed elevations are in dashed-dot, and evolving debris depths are in solid lines. The novel enhanced real two-phase model appropriately captures the emergence and propagation of complex frontal surge dynamics associated with the frontal ambient-drag and erosion. The surge is amplified with erosion. Simulations dynamically update the flowing materials, its two-phase rheologies, and phase interactions by incorporating the bed material into the flow.
\\[3mm]
In Fig. \ref{Fig_3} it is revealed that in connection to the erosion, the mass and momentum production or loss should essentially be included in the simulation. The panels, respectively, represent: {\bf \textcircled{a}} erosion with respective mass and momentum production, {\bf \textcircled{b}} physically incomplete erosion disregarding momentum production, {\bf \textcircled{c}} no mass but momentum production, which is equivalent to effectively reduced friction without mass production, and {\bf \textcircled{d}} no erosion at all. Only panels {\bf \textcircled{a}} and {\bf \textcircled{d}} are physically meaningful, representing the erosional scenario, and no erosion, respectively. These simulations reveal that momentum is lost in erosion, and when mass and momentum productions are appropriately incorporated into the flow dynamics, this ultimately adds momentum, and thus, the momentum production results in higher flow mobility as measured by the travel distance or the frontal position. As discussed earlier, such a phenomenon has often been observed in mass flows involving erosion (see, e.g., Iverson et al., 2011, de Haas et al., 2015). This clearly demonstrates that any conservative model that neglects the momentum changes (production, or loss) due to erosion is physically incorrect, and mathematically inconsistent. 
\\[3mm]
Furthermore, panel {\bf \textcircled{c}} shows unexpected flow mobility that resulted due to the invalid utilization of the momentum productions without the corresponding mass productions (result shown only until $t = 8$ s, as for larger time it exits the computational domain). This led to non-physical excess mobility because extra force is applied in the momentum balances without incorporating the eroded masses in the mass balances that would balance the momentum productions through the changes of momenta in the inertial parts of the momentum balances.

\section{Summary}

Two different erosion models are present in the literature - empirical and mechanical ones. However, both of them are (mostly) effectively single-phase. The mechanical erosion models involve the dynamical variables associated with the flow coupled to several physical parameters. Nevertheless, the existing mechanical erosion models may contain singularities, or result in non-physical behavior, for very low or high velocities. Here, we developed a two-phase, process-based, non-singular mechanical model for erosion rate for both the solid and fluid phases. The model is based on the jump in the momentum flux and enhances an existing general two-phase mass flow model (Pudasaini, 2012). The jump includes the contrasts in shear stress and momentum fluxes across the erodible substrate. At the interface the solid stress satisfies the Coulomb law, and the fluid stress follows the Chezy-type friction. The solid and fluid velocities on either sides of the singular surface are expressed in terms of the mean flow velocities. This introduces erosion drift coefficients. The net interfacial (solid) velocity is modelled by the shear velocity which introduces a further coefficient that transforms the shear stress to velocity. The singularities in the erosion rates have been systematically removed by developing mechanical closure models for the coefficients emerging in the erosion rates constructed here. This is based on the fact that effectively reduced frictional stress in erosion is equivalent to the momentum production. 
\\[3mm]
In contrast to existing models, the erosion drift equation clearly demonstrates that erosion cannot take place by setting the zero velocity of the eroded particle (or, molecule). We proved that if the basal erosion drift coefficient approaches unity then, the erosion rate increases rapidly as it happens for very light basal surface material. We further reveal the fact that when the basal substrate is weaker (or, stronger, in terms of density and volume fraction, or mass fraction component) as compared to the debris flow then, the eroded particle moves with relatively faster velocity (or, vice versa). These are natural conditions. For no density and volume fraction contrasts across the interface, results show that the particle at the bottom of the flow moves twice as fast as the velocity of the eroded particle on the other side of the interface (basal surface). In the most simple situation, for the plug flow, the basal drift coefficient becomes 1/2, a typical scale characterizing the erosion speed. Similarly, in this simple situation, the shear velocity factor is inversely proportional to the difference in the friction coefficients, difference in the buoyancy factors, and the solid volume fraction contrasts on either side of the erosion interface. Thus, the erosion process ceases as these differences tend to vanish.
\\[3mm]
The proposed erosion rate models explicitly show, that for erosion to take place, at least one of the four parameters or, the dynamical variables, must have a jump across the interface: the buoyancies, the densities, friction coefficients and the volume fractions. {Most existing erosion models are single-phase. However, it is crucial to consider two phase flows and two-phase erosion models. Because, considering the erosion rates or especially the reduced erosion rates, it appears, that one of the main dynamic variables driving the two phase erosion mechanism is the dynamically evolving solid (or, fluid) concentration. This characterizes the two-phase nature of the flow and erosion. The model presented here can also be applied to single phase flows when a friction (and/or density) jump across the interface is apparent in such flows.} It appears that even for basal material with higher friction than that for the moving mass, erosion may take place. Apparently, for solid (or, the solid-type bulk), erosion is not necessarily and directly dependent on the velocity but depends on the competition between the mechanical strength of the flowing debris and the bed material. However, for the fluid, the erosion rate is linearly proportional to the fluid flow velocity. We also explained how the present method can be applied to further develop other erosion rate models, including the bulk-type mixture models. We showed that mechanically deposition is the reverse process of erosion. We explained the situations on how the deposition can be triggered and how it is amplified with the applied forces, and how the evolving enhanced net basal shear stress overtakes the flow shear stress.  
\\[3mm]
 We proved that the erosion resulting in higher geophysical mass flow mobility can be described mechanically, and quantitatively when the mass and momentum productions are consistently and physically correctly included into the mass and momentum equations. We consider that erosion takes place only when the basal substrate is mechanically weaker than the flow itself. So, due to the weaker bed, as the effective resistance is reduced erosion takes place. The effectively reduced frictional stress, which is a positive quantity, is balanced by the system in terms of the momentum production. Furthermore, as the mass is added into the system, the gravity load immediately accelerates the total mass down the entire travel distance. This further enhances the flow mobility, because the erosion-induced added mass implies added potential energy. Altogether, this explains the higher mobility associated with the erosion in mass flow. With this, we solved the long standing dilemma of erosion related mobility of geophysical mass flows.
\\[3mm]
The model reveals some major aspects of the mechanics associated with erosion, entrainment and deposition. Simulations indicate that the model appropriately captures the emergence and propagation of complex flow dynamics associated with erosion, including the sharp frontal surge, and long tail with evolving basal surface. 
\\[3mm]
{\large \bf Acknowledgements:}  This work has been conducted as part of the international cooperation projects: ``Development of a GIS-based Open Source Simulation Tool for Modelling General Avalanche and Debris Flows over Natural Topography (avaflow)'' supported by the German Research Foundation (DFG, project number PU 386/3-1) and the Austrian Science Fund (FWF, project number I 1600-N30).

\end{document}